\documentclass[aip,
 amsmath,amssymb,
preprint,%
author-year,%
author-numerical
]{revtex4-2}

\usepackage{graphicx}
\usepackage{dcolumn}
\usepackage{bm}

\usepackage[utf8]{inputenc}
\usepackage[T1]{fontenc}
\usepackage{mathptmx}
\usepackage{etoolbox}
\usepackage{placeins}

\usepackage{xcolor} 
\usepackage{hyperref}
\hypersetup{
  colorlinks   = true,  
  urlcolor     = black, 
  linkcolor    = black, 
  citecolor    = black, 
  filecolor    = black  
}

\newcommand*{\rva}[1]{#1}
\newcommand*{\rvb}[1]{#1}

\newcommand{\comment}[1]{}

\newcommand{\s}[1]{\, {\rm #1}} 
\DeclareMathOperator*{\argmin}{{\rm arg \, min}}
\makeatletter
\def\@email#1#2{%
 \endgroup
 \patchcmd{\titleblock@produce}
  {\frontmatter@RRAPformat}
  {\frontmatter@RRAPformat{\produce@RRAP{*#1\href{mailto:#2}{#2}}}\frontmatter@RRAPformat}
  {}{}
}%
\makeatother

\begin{document}
\title[Learning Chaotic Dynamics with Neuromorphic Network Dynamics]{Learning Chaotic Dynamics with Neuromorphic Network Dynamics}
\author{Y. Xu}
 \affiliation{School of Physics, University of Sydney, Sydney, NSW 2006, Australia.}
\author{G.A. Gottwald}
\affiliation{ 
School of Mathematics and Statistics, University of Sydney, Sydney, NSW 2006, Australia
}
\author{Z. Kuncic}
\affiliation{
School of Physics, University of Sydney, Sydney, NSW 2006, Australia.
}
\affiliation{Centre for Complex Systems, University of Sydney, Sydney, NSW 2006, Australia.}
 \email[]{zdenka.kuncic@sydney.edu.au}
\date{\today}

\comment{ 	
This study provides valuable insights into how a physical neuromorphic network device can be optimised for learning complex dynamical systems solely through external input parameters.
Nonlinear and dynamical effects in the neuromorphic network are identified and investigated via simulating every component of the network, which allows for a deeper understanding of its inner workings.
This is a necessary step in the development of practical neuromorphic nano-devices, which may leverage their brain-like properties to achieve fast, cheap and energy efficient computations, via computing implicitly with the physics of the underlying neuromorphic system itself.
This study is also the first to successfully demonstrate autonomous prediction of a multivariate chaotic time series (specifically, the Lorenz63 attractor) with a neuromorphic nanowire network.
}
\begin{abstract}
This study investigates how dynamical systems may be learned and modelled with a neuromorphic network which is itself a dynamical system.
The neuromorphic network used in this study is based on a complex electrical circuit comprised of memristive elements that produce neuro--synaptic nonlinear responses to input electrical signals.
To determine how computation may be performed using the physics of the underlying system, the neuromorphic network was simulated and evaluated on autonomous prediction of a multivariate chaotic time series, implemented with a reservoir computing framework.
Through manipulating only input electrodes and voltages, optimal nonlinear dynamical responses were found when input voltages maximise the number of memristive components whose internal dynamics explore the entire dynamical range of the memristor model.
Increasing the network coverage with the input electrodes was found to suppress other nonlinear responses that are less conducive to learning.
These results provide valuable insights into how a physical neuromorphic network device can be feasibly optimised for learning complex dynamical systems using only external control parameters.
\end{abstract}

\maketitle

\section{Introduction}
Neuromorphic computing aims to achieve efficient computation by emulating the brain's powerful information processing abilities \cite{Christensen_2022,schuman2022opportunities,Song_2023, li2024brain,kudithipudi2025neuromorphic}.
Two broad approaches that have been successfully demonstrated are: (i) electrical circuits that emulate point neuron spike models and/or spike--based learning models \cite{maass1997networks,schuman2022opportunities, eshraghian2023training}; and (ii) nano--electronic materials with inherent neuro--synaptic dynamics \cite{ielmini2021brain,Song_2023,vahlBraininspired2024}. 
The latter includes resistive switching memory (memristive) materials \cite{chua2003memristor, strukov2008missing, di2009circuit, caravelli2018memristors}. 
Neuromorphic computing with memristors has been demonstrated with crossbar arrays, which can be used to perform physical matrix--vector operations to accelerate inference of artificial neural network models \cite{kim2012functional,li2019long}. 
Alternately, memristor crossbar arrays have also been used to implement reservoir computing (RC) \cite{moonTemporal2019,Yan2024}. 
Neuromorphic RC with memristive systems is a particularly promising approach given that electronic reservoirs with neuromorphic dynamics closely resemble the brain's physical reservoir \cite{Liang_2024,Enel_2016}. 
However, a drawback of memristor crossbars for neuromorphic RC is their regular structure and limited scale \cite{Larremore2011,Milano_etal_2022,Damicelli2022,Kumar_etal_2022,Millan2025,Jain2025}.

This study focuses on neuromorphic RC using memristive nanowire networks. 
These are complex, self--organised electrical circuits comprised of axon-like nanowires and synapse-like memristive junctions between overlapping nanowires, where conductance changes --- the physical realisation of synaptic weights --- are triggered by migration of \rva{ions} similar to the transmission of neurotransmitter molecules, constrained by physical equations of state and conservation laws \cite{kuncicNeuromorphic2021}.
Memristive nanowire networks exhibit emergent brain--like dynamics, such as dynamical phase transitions (continuous and discontinuous) and synchronisation. Previous studies have shown that different dynamical regimes may be exploited for learning \cite{hochstetterAvalanches2021, zhuInformation2021, baccettiErgodicity2024, pilati2024emerging}. 
As a consequence of their self--organisation (i.e. their ability to acquire non--trivial structure--function properties), a single voltage pulse may have cascading effects that evolve all memristive elements in the neuromorphic network, in contrast to digital implementations of artificial neural networks, where large numbers of transistors must be individually addressed to update a similar number of network weights \cite{jaeger2023toward}.

Nanowire networks naturally self--organise into a Recurrent Neural Network (RNN) structure, with a functional connectivity that is highly suitable for RC.
In standard algorithmic RC, the reservoir is typically a fixed random RNN of which each node evolves according to a mathematical activation function, and only a single linear output layer is trained with a computationally efficient linear regressor or classifier \cite{lukoseviciusReservoir2009}.
In this way, RC approaches can achieve excellent performance in various learning tasks involving temporal data \cite{pathakModelFree2018, jaegerAdaptive2002, fernandoPattern2003}, with training times that are orders of magnitude shorter than standard machine learning approaches. 
Physical RC, however, may operate differently because of the dynamics of the physical reservoir.
In the case of neuromorphic RC with memristive nanowire networks, all the network dynamics are driven by the memristive edge junctions and are constrained by physics \cite{Milano2023strategies,kuncicNeuromorphic2020a,Fu2020,zhuInformation2021,hochstetterAvalanches2021,Milano_etal_2022,Daniels_etal_2022,Milano_2022,ijcnn_unpublished,Monaghan_etal_2025}.

To further investigate how memristive nanowire networks operate as physical reservoirs for neuromorphic RC, 
we present a simulation study focusing on a specific application: autonomous prediction of a multivariate chaotic time series, the well--known Lorenz63 system \cite{lorenzDeterministic1963}. 
Prediction of the Lorenz63 system has been previously demonstrated with several different RC and related approaches \cite{kosterDatainformed2023,akiyamaComputational2022,choi20243d, shahi2022prediction,gilpinModel2023, mandal2025,Brunton2016,Kan2021,mandal2024choice}. 
To the best of our knowledge, this study is the first to demonstrate autonomous prediction of the Lorenz63 attractor with physical neuromorphic RC. 
Understanding and predicting the behaviour of complex nonlinear dynamical systems is crucial for many scientific disciplines, including physics, climatology, neuroscience and biology, and for real--world applications such as financial markets and social networks \cite{dos2025reservoir, strogartz1994nonlinear,strogatz2001exploring}.
Chaotic dynamical systems, in particular, are notoriously difficult to forecast due to their high sensitivity to initial conditions and highly nonlinear nature, resulting in aperiodic and categorically unpredictable dynamics. 
As such, the task of forecasting chaotic dynamics provides a challenging testing ground for exploring the limits of the capabilities of neuromorphic networks, and for testing the hypothesis that neuromorphic physical reservoirs can predict dynamical systems using the inherent physics of the neuromorphic system itself~\cite{Legenstein-Maass_2007,Dambre_etal_2012,markovic2020physics,Kan2021,jaeger2023toward, caravelli2025self}.

This study primarily aims to gain deeper insights into the inner workings of memristive nanowire networks for learning via neuromorphic RC.
No attempt is made to achieve state-of-the-art prediction results \rva{for the Lorenz system}, such as those already achieved with random feature models\cite{mandal2025}, Long-Short Term Memory (LSTM) networks \cite{vlachas2020backpropagation} or algorithmic RC models \cite{platt2022systematic}. 
Rather our study provides the first proof of concept that analog devices consisting of a physical neuromorphic network are capable of \rva{learning the Lorenz63 attractor}, through controlling the network via only the external input parameters.
The forecast performance achieved in our study for a network with $2,000$ nodes is far below those achieved by digital reservoir computers of the same size. 
However, as we show, performance increases with network size, and physical neuromorphic networks can readily scale to millions of nodes \cite{diaz-alvarezEmergent2019,sillin2013theoretical}; 
such network sizes are out of reach for digital reservoir computers.
Also considering optimisation of the computational framework and the network itself (which is beyond the scope of this study), physical neuromorphic devices have great potential to outperform digital computers, with significantly faster and more energy efficient performance \cite{Christensen_2022,Song_2023,kudithipudi2025neuromorphic}. 

The remainder of this article is organised as follows.
Sect.~\ref{sec:method} describes the methods used in the simulation study, including the RC setup and memristive network model. 
The results in Sec.~\ref{sec:results} start with analysis of the chaotic time series prediction task (in Sec.~\ref{sec:lorenz}), then presents and discusses all sources of nonlinear dynamical features, on the global network level in Sec.~\ref{sec:nonlin}, then local memristive edge level and their interactions in Sec.~\ref{sec:transdynam} and Sec.~\ref{sec:node_inter}. 
This study concludes with Sec.~\ref{sec:conclusions}, which shines light on the future works needed to fully optimise this neuromorphic network for modelling dynamical systems. 

\section{Methods}\label{sec:method}

A neuromorphic reservoir was constructed using a physically--motivated model of a complex electrical circuit comprised of nonlinear resistive switching memory circuit elements known as memristors \cite{kuncicNeuromorphic2021, kuncicNeuromorphic2020a,Fu2020,zhuInformation2021, hochstetterAvalanches2021}. 
Given some input signal $\mathbf{u}(t) \in \mathbb{R}^{N_{\rm u}}$, the neuromorphic network is used to autonomously predict a dynamical time series $\mathbf{y}(t)\in \mathbb{R}^{N_{\rm y}}$, by computing the estimate $\mathbf{\hat{y}}(t)\in \mathbb{R}^{N_{\rm y}}$ of $\mathbf{y}(t)$.
This is performed using a reservoir computing (RC) framework in which reservoir input values $\mathbf{r}_{\rm in}(t)$ are coupled to the input signal $\mathbf{u}(t)$ via a linear input layer with weight matrix $W_{\rm in} \in \mathbb{R}^{N_{\rm in} \times N_{\rm u}}$ and an input bias vector $\mathbf{b}_{\rm in} \in \mathbb{R}^{N_{\rm in}}$, such that 
\begin{equation}
	\mathbf{r}_{\rm in}(t)= W_{\rm in} \; \mathbf{u}(t) + \mathbf{b}_{\rm in}.\label{eq:rin}
\end{equation}
with $W_{\rm in}$ and $\mathbf{b}_{\rm in}$ entries uniformly randomly sampled on intervals $[-w, w]$ and $[-b, -b/2]\cup [b/2, b]$, respectively (the choice of the bias sampling is explained in Sec.~\ref{sec:node_inter}). We choose $w=\alpha$, $b=5\alpha$ with $\alpha = 0.2\s{V}$.
The $\alpha$ variable effectively serves as the input voltage scaling parameter which converts the input signals into appropriate voltages.

\begin{figure*}
\begin{center}
    \includegraphics[width=0.75\linewidth]{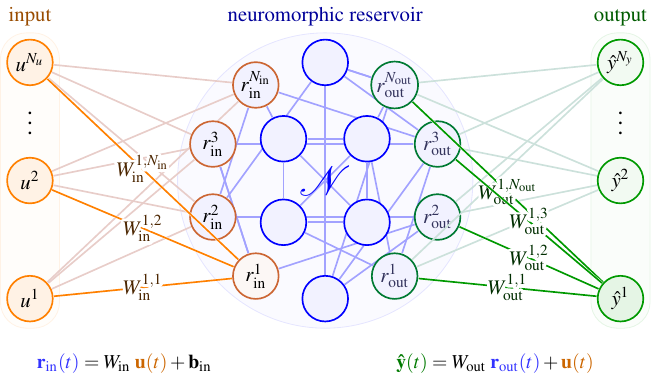}
\end{center}
\caption{
	Schematic of the dynamic neuromorphic reservoir computer. 
	A signal vector $\mathbf{u}(t)$, weighted by a fixed random $W_{\rm in}$ and scaled by a constant voltage $\alpha$, is input into the neuromorphic reservoir $\cal{N}$ via selected input nodes. 
	Voltage signals $\mathbf{r}_{\rm in}(t)$ (which also include random bias values $\mathbf{b}_{\rm in}$) are mapped into a higher-dimensional dynamical feature space which is sampled from other nodes $\mathbf{r}_{\rm out}(t)$. 
	Only the output weight matrix $W_{\rm out}$ is trained to learn estimates $\mathbf{\hat{y}(t)}$. 
}
\label{fig:rc_schematic}
\end{figure*}

A graph representation of the neuromorphic network ${\cal N}$ consisting of $N$ nodes is abstracted as the reservoir (see Fig.~\ref{fig:rc_schematic}), a high dimensional dynamical system which nonlinearly transforms a temporal signal $\mathbf{r}_{\rm in}(t) \in \mathbb{R}^{N_{\rm in}}$
into readout values $\mathbf{r}_{\rm out}(t) \in \mathbb{R}^{N_{\rm out}}$, with $N_{\rm out} \geq N_{\rm in}$.
Only a subset $N_{\rm in}<N$ of all available nodes are used as input nodes, 
and $N_{\rm out}<N$ is the number of voltage readout nodes. 
In a machine learning context, these readouts $\mathbf{r}_{\rm out}(t)$ serve as high-dimensional dynamical feature embeddings that can be used to learn the nonlinear dynamics of the input data \cite{Carroll2021}. 
The neuromorphic network model is described further in Sec.~\ref{sec:nwn}.

The dynamical features $\mathbf{r}_{\rm out}(t)$ generated by ${\cal N}$ are linearly coupled to an external output layer with a weight matrix $W_{\rm out} \in \mathbb{R}^{N_{\rm y} \times N_{\rm out}}$ that is trained to learn the dynamics of $\mathbf{u}(t)$ by optimising output estimates $\mathbf{\hat{y}}(t)$ according to 
\begin{equation}
	\mathbf{\hat{y}}(t) = W_{\rm out}\; \mathbf{r}_{\rm out}(t)+ \mathbf{u}(t),\label{eq:yhat}
\end{equation}
which includes components $\mathbf{u}(t)$ that serve a similar role to skip connections in residual networks \cite{heDeep2015a}.

During training, the input signal is $\mathbf{u}(t) = \mathbf{y}(t-\Delta t)$, and
$W_{\rm out}$ is trained using ridge regression, given by
\begin{equation}
    W_{\rm out} = \argmin_{W_{\rm out}} \left(||W_{\rm out}\mathbf{r}_{\rm out}(t) - \Delta \mathbf{y} (t)||_{F}^2 + \gamma ||W_{\rm out}||^2_{F}\right),
\end{equation}
where $\Delta \mathbf{y}(t) = \mathbf{y}(t)-\mathbf{y}(t-\Delta t)$,
with a Tikhonov regularisation parameter value of $\gamma=10^{-6}$, and $||\cdot||_F $ denotes the Frobenius norm.

To perform dynamical system forecasting, the current estimate $\mathbf{\hat{y}}(t)$ at one time step ahead is predicted using the previous estimate $\mathbf{\hat{y}}(t-\Delta t)$ as input (i.e. $\mathbf{u}(t) = \mathbf{\hat{y}}(t-\Delta t)$).
Substituting into Eq.~\eqref{eq:yhat}, it can be seen that during prediction
\begin{equation}
	\frac{d\mathbf{\hat{y}}}{dt} \approx \frac{\mathbf{\hat{y}}(t)-\mathbf{\hat{y}}(t-\Delta t)}{\Delta t} = \frac{1}{\Delta t} W_{\rm out}\; \mathbf{r}_{\rm out}(t).
\end{equation}
Hence, with skip connections, the neuromorphic reservoir learns an Euler discretisation with fixed sample time $\Delta t$ of the dynamical equation.

Here we use the neuromorphic RC framework to perform autonomous prediction of the Lorenz63 system\cite{lorenzDeterministic1963}, given by 
\begin{align}
\begin{split}
   	\frac{d\tilde{y}_1}{dt} &= \rho_1 (\tilde{y}_2-\tilde{y}_1), \\ 
	\frac{d\tilde{y}_2}{dt} &= \tilde{y}_1(\rho_2 - \tilde{y}_3) - \tilde{y}_2 ,  \\ 	
	\frac{d\tilde{y}_3}{dt} &= \tilde{y}_1 \tilde{y}_2 - \rho_3 \tilde{y}_3. 
\end{split}\label{eqn:Lorenz63}  
\end{align}

The constants were set to the standard values of $\rho_1 = 10$, $\rho_2 = 28$ and $\rho_3 = 8/3$ that generate chaotic dynamics.
The normalised Lorenz values is recorded at equidistant sampling times $\Delta t=0.005$, and a training length of $135$ Lyapunov times was chosen to ensure a low training error (less than $0.01$ normalised root mean square error). 
The sampling time $\Delta t$ is used as the discretisation time step of our network simulations. 
All three discretised variables of the Lorenz63 system in Eq.~\eqref{eqn:Lorenz63} were provided as input after being normalised to a mean of $0$ and standard deviation of $1$, i.e. the $i$th element of $\mathbf{y}(t)$ is
\begin{equation}
    y_i(t) = \frac{\tilde{y}_i(t) - \langle \tilde{y}_i(t) \rangle}{\sigma_{\tilde{y}_i}},
\end{equation}
where $\langle \cdot \rangle$ denotes the time average, and $\sigma_{\tilde{y}_i}$ is the standard deviation (across all training time) of the $i$th component of the Lorenz system.

For quantifying prediction accuracy, the forecast time $t_f$ was determined as the longest time such that the relative forecast error $\mathcal{E}_f(t_f)$ is less than a threshold $\theta$:
\begin{equation}
    t_f = \max_t{(\lambda_{\rm max} t|\mathcal{E}_f(t)\leq\theta)}
\  \text{ with } \ 
    \mathcal{E}_f(t) = \frac{||\mathbf{y}(t) - \mathbf{\hat{y}}(t)||^2}{\langle||\mathbf{y} - \langle \mathbf{y} \rangle||^2\rangle},
    \label{eqn:err}
\end{equation}
where  $||\cdot||$ denotes the Euclidean norm.
A threshold of $\theta=0.4$ was chosen to be comparable with related studies \cite{kosterDatainformed2023,akiyamaComputational2022}.
The forecast time $t_f$ is measured in units of the Lyapunov time $\lambda_{\rm max}^{-1}$, where the largest Lyapunov exponent of the Lorenz63 system is $\lambda_{\rm max}=0.91$. 

\subsection{Neuromorphic Network Model}\label{sec:nwn}
The neuromorphic network $\cal N$ is comprised of a complex memristive circuit as a model of self--organised nanowire networks \rvb{on a 2D plane}, which exhibit neuro--synaptic dynamics under electrical stimulation \cite{kuncicNeuromorphic2021}.
Due to its unique network topology, the neuromorphic networks were generated by modelling the bottom--up self--assembly process of the physical nanowire network; see Ref.~\cite{loeffler2020topological} for detailed network generation and network topology analysis. 

\begin{figure}
    \centering
    \includegraphics[width=0.99\linewidth]{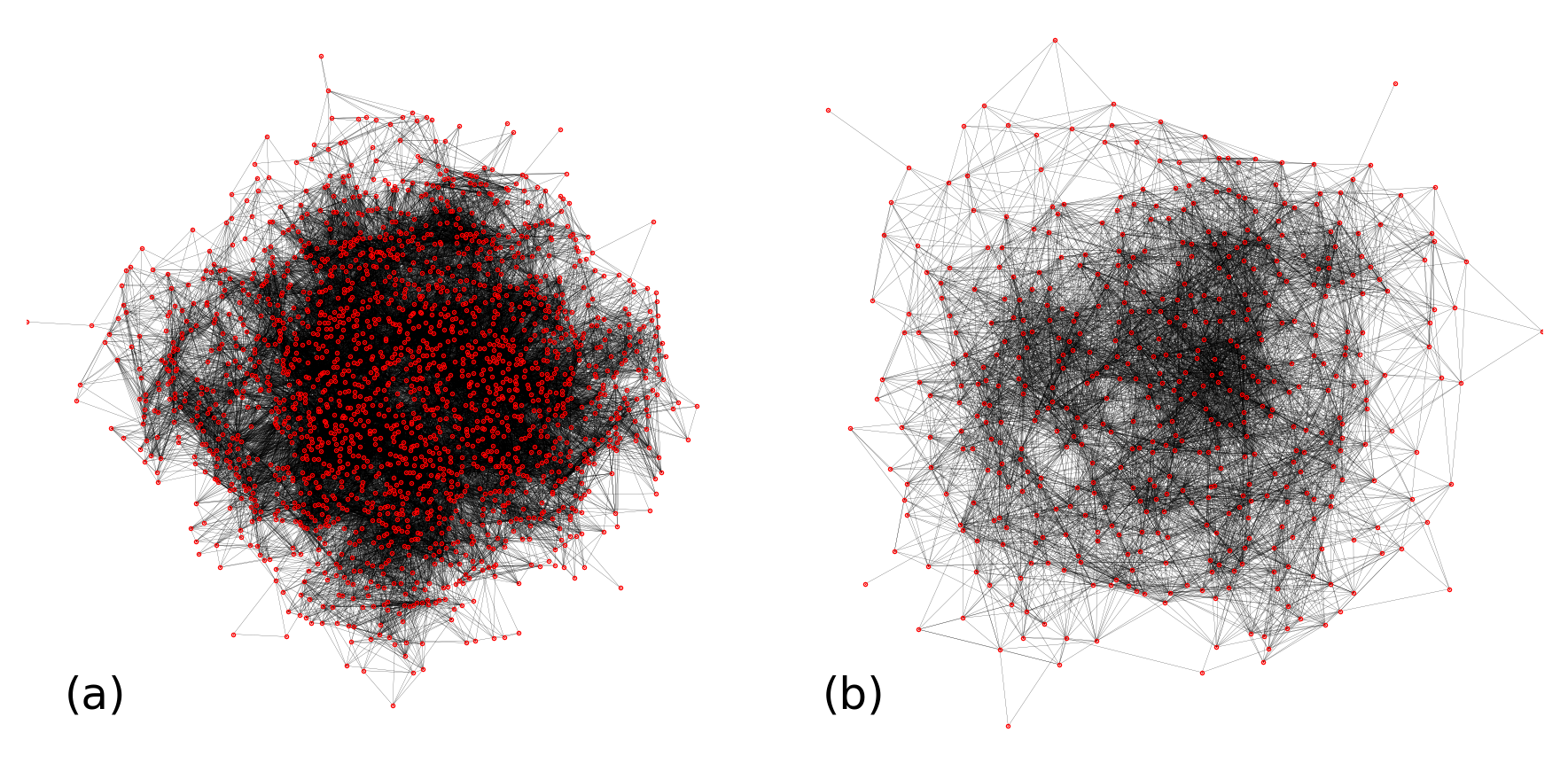}
	\caption{
		Graph representations of simulated neuromorphic networks used in this study: (a) network with $2,000$ nodes and $47,946$ edges; (b) network with $500$ nodes and $9,905$ edges. \rvb{The corresponding degree distributions are shown in Appendix Fig.~\ref{fig:degree_dist}}.}
	\label{fig:networks}
\end{figure}

The neuromorphic network $\cal N$ is abstracted as a graph representation, with nodes representing nanowires and edges representing memristors. 
\rvb{The resulting network is undirected and connected (no isolated subgraphs).}
A network of $2,000$ nodes and $47,946$ edges, alongside another network of $500$ nodes and $9,905$ edges were used, with $5$\% of all nodes serving as inputs and up to $90$\% of all other nodes serving as readout nodes. 
Their graph representations are illustrated in Fig.~\ref{fig:networks}.
\rvb{Both networks have an average degree of $\simeq 40$, and a negatively skewed log-normal degree distribution (see Fig.~\ref{fig:degree_dist} in Appendix \ref{app:supp}). }

\begin{figure}
	\centering
	\includegraphics[width=\linewidth]{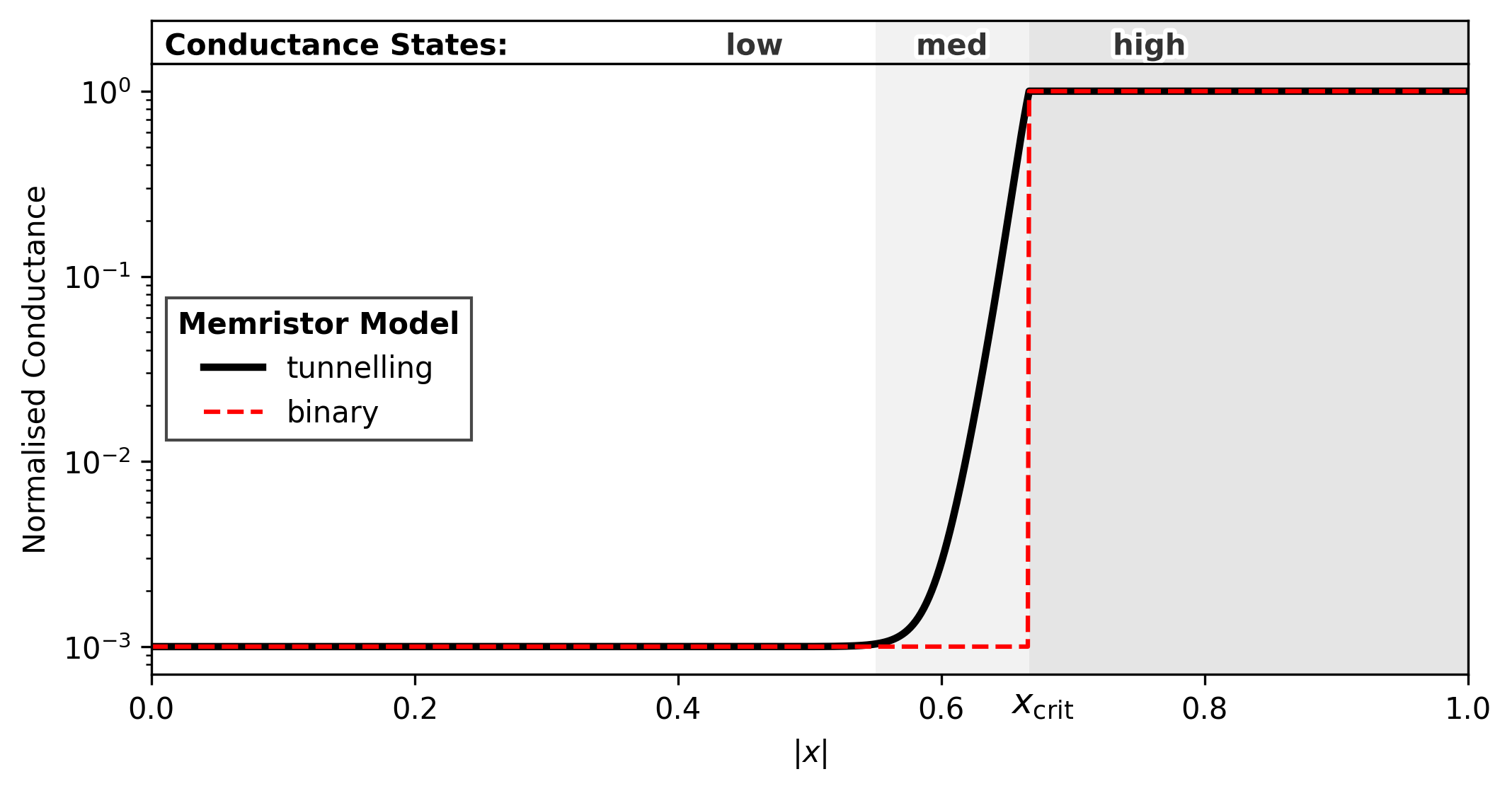}
	\caption{Memristive edge normalised conductance ($g(x)/\max_x[g(x)]$) as a function of the internal state parameter $x$, for both the tunnelling memristor model (black) and the binary model (dotted red).
	The low, medium and high conductance states are indicated. }
	\label{fig:conductance}
\end{figure}

The graph Laplacian of $\cal N$ is used to define the linear system of equations corresponding to Kirchhoff’s and Ohm's laws for electrical circuits. 
For an arbitrary, non--trivial network with $N$ nanowires and $N_{\rm in}$ contact electrodes, for every specified time-step, the node voltages $\mathbf{r}$ and electrode currents $\mathbf{i}_{\rm in}$ are calculated by solving the following system of equations in response to the external input signals $\mathbf{r}_{\rm in}$~\cite{ho1975modified,dorfler2018electrical}
\begin{equation}
    L
	\left[\begin{array}{c}\mathbf{r}\\ \mathbf{i}_{\rm in}\end{array}\right]
	=
	\left[\begin{array}{c} \mathbf{0} \\ \mathbf{r}_{\rm in} \end{array}\right],
	\label{eq:NL}
\end{equation}
where $L$ in its block matrix representation is 
\begin{equation}
    L=
    	\left[\;\begin{array}{@{}c|c@{}}
		\mathcal{L} & C^{T} \\
		\hline
		C & 0_{N_{\rm in}}
	\end{array}\;\right].\label{eq:L}
\end{equation}
Here, $\mathcal{L} = D - W$ is the graph Laplacian of size $N\times N$, $W$ is the weighted adjacency matrix of the network with each of its element given by
	$
	W_{ij} = g_{ij} A_{ij},
	$
	where $A$ is the adjacency matrix of the network, and $g_{ij}$ is the conductance of the edge (junction) connecting the $i$th and $j$th nodes (nanowires), which evolve dynamically with respect to past voltage values (described formally in Sec. \ref{sec:memristive_edges}).
	The $N\times N$ matrix $D$ is the weighted degree matrix given by 
	$D = \text{diag} (d_i)$, $d_i = \sum_{l=1}^{N} W_{il}$.
	$C$ is a $N_{\rm in} \times N$ matrix, with $C_{ij}=1$ for electrodes with index $i$ that are assigned on the node with index $j$, and $C_{ij}=0$ otherwise.
	$0_{N_{\rm in}}$ is a zero matrix of size $N_{\rm in} \times N_{\rm in}$, and
	$\mathbf{0}$ is a $1\times N$ zero vector.
	$\mathbf{r}$ is a $1 \times N$ vector encoding the voltage of all $N$ nanowire nodes. 

The voltage $v_{ij}$ across the edge connecting the $i$th and $j$th node is $v_{ij} = r_i - r_j$, where $r_i$ and $r_j$ are the node voltages at the $i$th and $j$th index, respectively.
Once the node voltages are solved, an internal state parameter $x_{ij}$ is updated for each junction with a memristor model. 
The model is then used to calculate the conductance $g_{ij}$ at each memristive junction, which can be used to update $\mathcal{L}$ for the next timestep.
\subsection{Memristor Model}\label{sec:memristive_edges}
We use a threshold memristor model~\cite{hochstetterAvalanches2021,kuncicNeuromorphic2020a,zhuInformation2021}, given by the following expression for a single memristive junction with voltage $v(t)$ (indices omitted for clarity),
\begin{equation}
	\frac{dx}{dt} = \left\{
	\begin{array}{ll}
		\eta(|v(t)|-V_{\text{set}}) \;
        \text{sgn}(v(t))
        & \quad |v(t)|>V_{\text{set}} \\
		0 & \quad V_{\text{reset}} \leq |v(t)| \leq V_{\text{set}}\\
		\eta q (|v(t)| - V_{\text{reset}}) 
        \;\text{sgn} (x(t))
        & \quad V_{\text{reset}} > |v(t)|\\
		0 & \quad |x|\geq 1
	\end{array}\right.\label{eq:dx}
\end{equation}
\rva{The physico-chemical processes responsible for memristive switching include electro-chemical metallisation (ECM) and electron tunnelling. Here, only electron tunnelling is explicitly modelled to update the dimensionless state variable $x(t)$. Other models have been proposed that capture ECM, but not electron tunnelling~\cite{Miranda_etal_2020,Milano_etal_2022_materia}.}
Constant $q$ controls the decay of memory of the memristor, and $\eta$ controls the time scale of the memristor relative to the input signal.

The memristive junction conductance $g(x)$ only depends on $x(t)$ and is expressed as
\begin{equation}
	g(x) = \frac{1}{R_t(x)+R_{\rm on} + \frac{R_{\rm on}^2}{R_{\rm off} - R_{\rm on}}} + \frac{1}{R_{\rm off}},\label{eq:g}
\end{equation}
where the tunnelling conductance $G_t(x) = R_t(x)^{-1}$ assumes the low voltage Simmon's formula \cite{Simmons1963GeneralizedFF}:
\begin{equation}
	G_t(x) = \frac{\zeta_0}{s}  \exp{\left(\zeta_1 s\right)},
    \label{eq:simmons}
\end{equation}
with 
\begin{equation}
	s = \max{\left[ (x_{\rm crit} - |x|) \frac{s_{\max}}{x_{\rm crit}}, 0\right]}. 
\end{equation}
All parameter values not explicitly stated here are listed in Appendix \ref{app:sim}.

We further consider a simpler binary memristor model where the junction conductance $g$ is either $R_{\rm on}^{-1}$ or $R_{\rm off}^{-1}$, which is effectively equivalent to the above memristor model without electron tunnelling transport:
\begin{equation}
	g(x) = \left\{
	\begin{array}{ll}
	 	R_{\rm on}^{-1}& \quad |x|>x_{\rm crit}, \\
		R_{\rm off}^{-1}& \quad |x|\leq x_{\rm crit}.\\
	\end{array}\right.\label{eq:binaryG}
\end{equation}
The conductance profiles of both memristor models are illustrated in Fig.~\ref{fig:conductance}.

We remark that the nontrivial temporal evolution of the internal state variable in Eq.~\eqref{eq:dx} \rvb{parametrises internal nonlinear conductance changes via Eq.~\eqref{eq:g}.}  
If the state variables do not evolve (i.e. ${\dot{x}}_{ij}=0$ for all memristive junctions), the neuromorphic network reduces effectively to a network of linear resistors.

Further details of the neuromorphic network and memristor models can be found in Refs.~\cite{zhuInformation2021, hochstetterAvalanches2021,baccettiErgodicity2024}.
\subsection{Network Dynamics}
\begin{figure*}
	\centering
	\includegraphics[width=\linewidth]{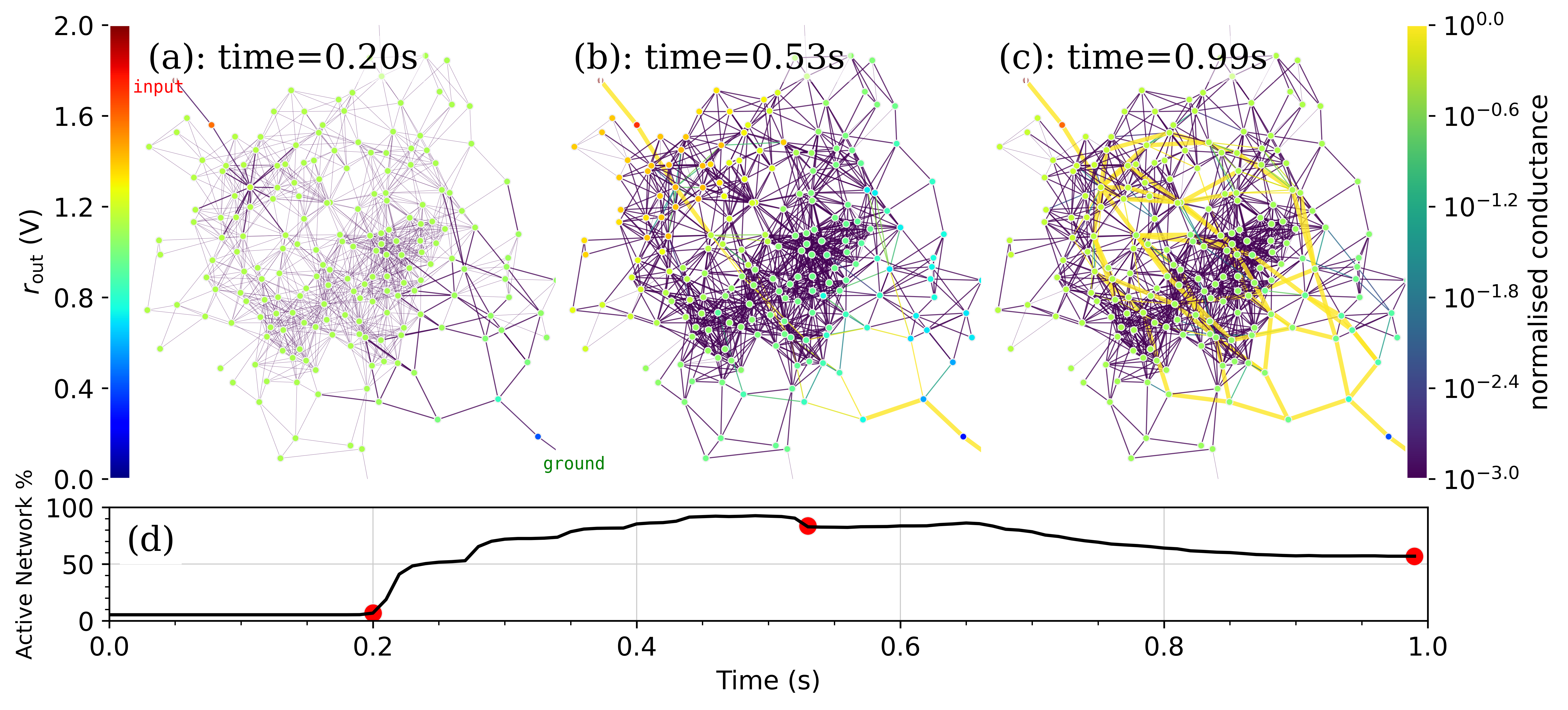}
	\caption{Graph visualisations of a simulated $200$-node, $1,213$-edge neuromorphic network showing dynamic connectivity in response to a constant DC voltage bias of $2\s{V}$. Snapshots are shown at times (a) $0.2$\,s, (b) $0.5$\,s and (c) $1.0$\,s, with the normalised conductance on edges, $g(x)/\max_x[g(x)]$, indicated by the right colourbar and the thickness of the respective edges, the node voltages $\mathbf{r}_{\rm{out}}$ indicated by the colourbar on the left; (d) corresponding percentage of the network that is active with edge voltage $v_{ij}\geq V_{\rm set}$.}
	\label{fig:snap1}
\end{figure*}

\rva{The neuromorphic network node dynamics are driven by the memristive dynamics described above.}
An example of a network's dynamical behaviour is presented in Fig.~\ref{fig:snap1}, which shows snapshots of conductance changes for different times as a voltage bias is applied to one node. 
As voltage is distributed to nodes according to their connectivity, edge conductances begin to evolve forming localised branches (Fig.~\ref{fig:snap1}(a)) that drive further node voltage redistribution, with the formation of the first high conductance path (Fig.~\ref{fig:snap1}(b), yellow) between the input and the ground node in the electrical circuit. 
As the input voltage persists, more parallel conductance paths form (Fig.~\ref{fig:snap1}(c)).
\rva{These paths reset under reverse voltage polarity and/or via stochastic fluctuations associated with atomic nanofilament formation and decay \cite{Milano_etal_2025}, which can also affect the path setting process. We do not model stochastic fluctuations here to preserve interpretability of results.}

For arbitrary input voltage signals, the network is only dynamically active in certain parts at different times due to its connectivity.
As shown in Fig.~\ref{fig:snap1}(d), network activity (defined by edge conductance changes when voltage difference are above a threshold set value $V_{\rm set}$) gradually increases, reaching a maximum of around $90$\% before decreasing towards $50$\%.
After $0.5$\,s, conductance changes slow down in tandem with reduced voltage differences between nodes.
With multiple input nodes and more varied, bipolar input signals, the network's behaviour follows similar dynamic patterns, but with richer complexity in its response. 
Recurrent edge--node interactions enable more nonlinear coupling between inputs and readouts, which may be advantageous for tasks involving multivariate time series with complex dynamics, such as the Lorenz system. 

\subsection{Dynamic Reservoir Computing}\label{sec:dynamicRC}
Due to the presence of memristive edge dynamics which induces complex network dynamics (cf. Fig.~\ref{fig:snap1}), the neuromorphic network behaves very differently when compared with classical reservoirs used in standard RC approaches.
In conventional RC, edges are random and fixed, and dynamics are introduced solely on the nodes with a prescribed nonlinear activation function $\sigma$ (such as tanh or ReLU) according to 
\begin{equation}
\mathbf{r}(t+\Delta t) 
= \sigma \left( W\mathbf{r}(t) + W_{\rm in} \mathbf{u}(t)\right), \label{eq:rc}
\end{equation}
where $W$ is the static reservoir network weight matrix. 
In contrast, for the neuromorphic reservoirs considered here, Eq.~\eqref{eq:rc} is replaced by 
\begin{equation}
    \mathbf{r}(t+\Delta t) 
= M W_{\rm in} \mathbf{u}(t),
\label{eq:M}
\end{equation}
where $M(t;\mathbf{r}(t), \mathbf{r}(t-\Delta t),... \mathbf{r}(0); \mathbf{u}(t), \mathbf{u}(t-\Delta t),...,\mathbf{u}(0))$ is a submatrix of $L^{-1}$,
where $L$ is the block matrix in Eq.~\eqref{eq:L}.
$M$ has size $N \times N_{\rm in}$, such that its elements $M_{ij} = L^{-1}_{ik}$, where $k=j+N$ for $i \in [1,N]$, $j\in [1,N_{\rm in}]$.
$M$ is updated at every timestep from the previous $M(t-\Delta t)$ and current input $\mathbf{u}(t)$. 
Hence, $M$ is dependent on all previous input signals and internal voltage values and encodes all conductance weight changes, via network connectivity and Kirchhoff's laws. 

In conventional reservoirs, all reservoir nodes are linearly combined to generate the output signal. \rvb{However in the physical neuromorphic reservoirs considered here,} due to the limited number of electrodes used in a physical device, only a subset of all reservoir node voltages, the so called readout nodes, is used to generate the output signal.
Another key difference when compared to conventional RCs is the presence of adaptive weights.
Although the dynamically evolving conductance values on edges cannot be interpreted as neural network weights, since the neuromorphic network circuitry induces \rva{network} dynamics instead of a matrix multiplication used in abstract neural networks, they serve a similar role as network weights. 
Indeed, the most important distinction is the different sources of nonlinearity: nonlinear effects in a conventional reservoir stem from the \rvb{mathematical} activation function imposed on its nodes, whereas neuromorphic networks \rvb{in this study derive their nonlinear effects from their internal physical dynamics and heterogeneous connectivity, which influences RC performance~\cite{Loeffler2021,ijcnn_unpublished}}.

\section{Results \& Discussion} 
\label{sec:results}

The results presented below aim to demonstrate how a neuromorphic network works as a physical reservoir and how it is implemented for dynamic RC, using the specific example of autonomous prediction of the Lorenz63 attractor.
Both short--term and long--term predictions are presented and analysed, and the influence of external global parameters (input voltage scaling and bias) on forecasting performance is investigated using input--output mappings and a dynamics measure. 
Nonlinear dynamical features produced by the neuromorphic network are closely examined to determine which features are learned by the auto--regressor used for prediction. How the desirable nonlinear features can be promoted by the set--up of the neuromorphic reservoir is investigated by studying electrode node connectivity.


\subsection{Lorenz System Prediction}\label{sec:lorenz}

\begin{figure}
	\centering
	\includegraphics[width=0.99\linewidth]{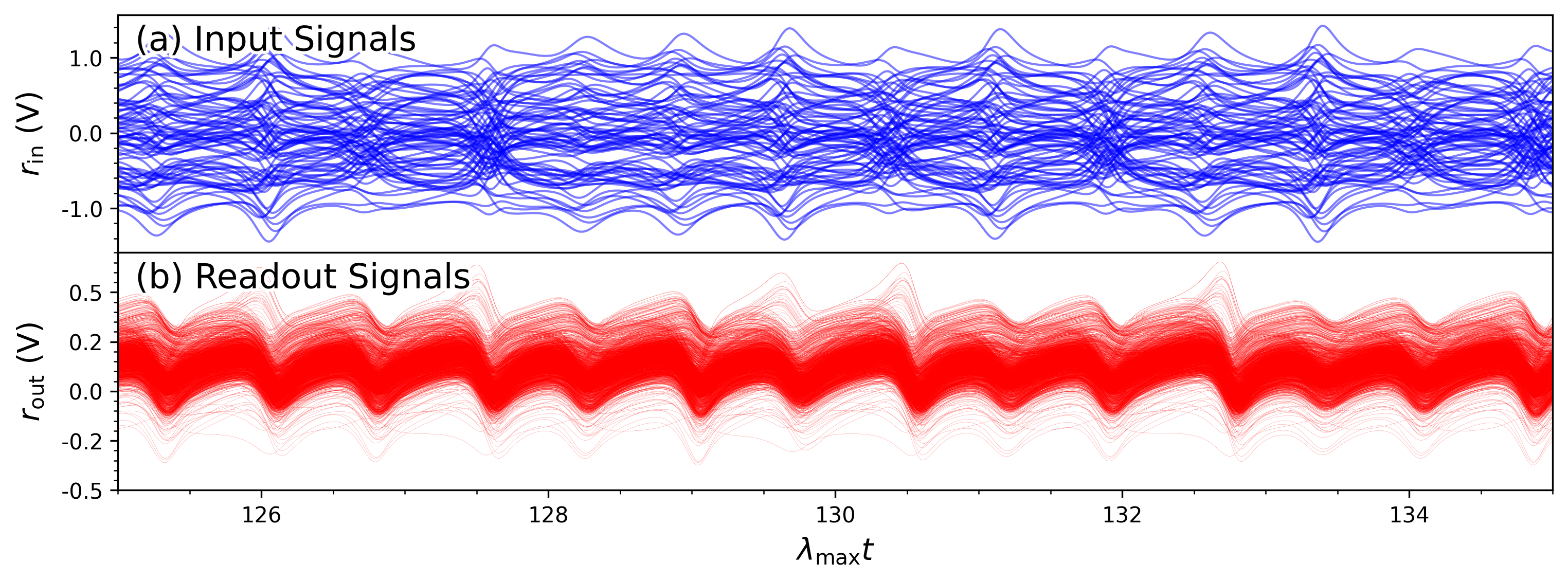}
	\caption{Input--output mapping of a neuromorphic network with $2,000$ nodes and $47,946$ edges. (a) Input signals $\mathbf{r}_{\rm in}(t)$ (blue) constructed from a linear combination of the Lorenz system (cf. Eq.~\eqref{eq:rin}) and delivered to $80$ randomly selected nodes; (b) readout signals $\mathbf{r}_{\rm out}(t)$ (red) from $1,919$ randomly selected nodes (distinct from the input nodes). 
	Time duration is the last 10 Lyapunov times of training. }
	\label{fig:in_out}
\end{figure}

\begin{figure}
	\centering
	\includegraphics[width=0.99\linewidth]{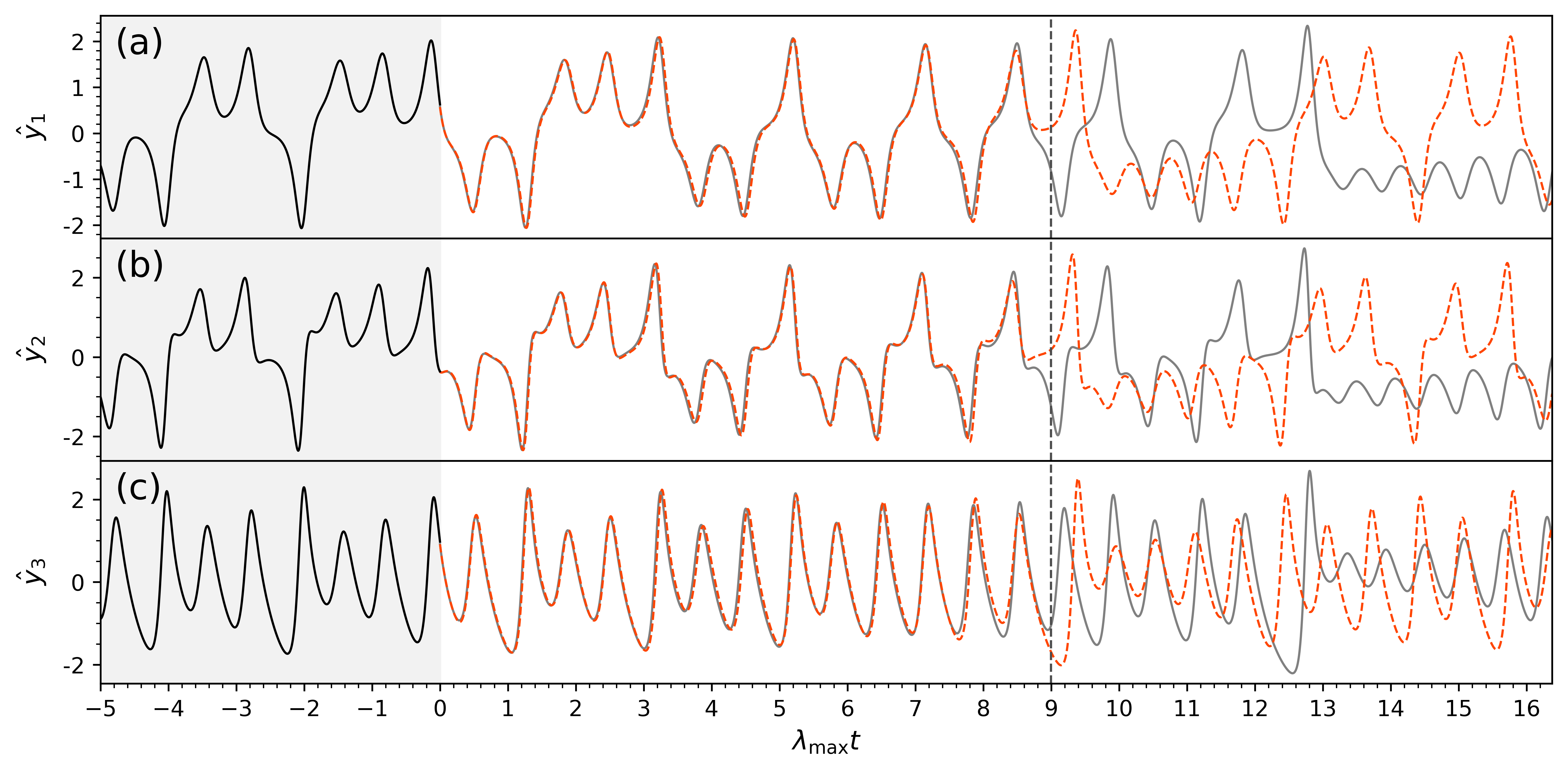}
	\caption{Example autonomous prediction of the Lorenz63 system \eqref{eqn:Lorenz63} using a neuromorphic network. (a), (b) and (c) respectively show the individual $\hat{y}_1(t)$, $\hat{y}_2(t)$ and $\hat{y}_3(t)$ values for the predicted signals (orange dashes) and the normalised true Lorenz signals (grey curves). The vertical grey dashed lines indicate the short-term prediction length, defined by Eq.~\eqref{eqn:err}, as around $9.0$ Lyapunov times. 
    \rvb{The shaded grey region shows the last 5 Lyapunov times of training for a total duration of $\simeq 135$ Lyapunov times, with NRMSE of $0.0012$. Autonomous prediction commences directly after training at $t=0$.}
    Data extracted from the same simulation as in Fig.~\ref{fig:in_out}.}
	\label{fig:pred1}
\end{figure}

\begin{figure}
	\centering
	\includegraphics[width=\linewidth]{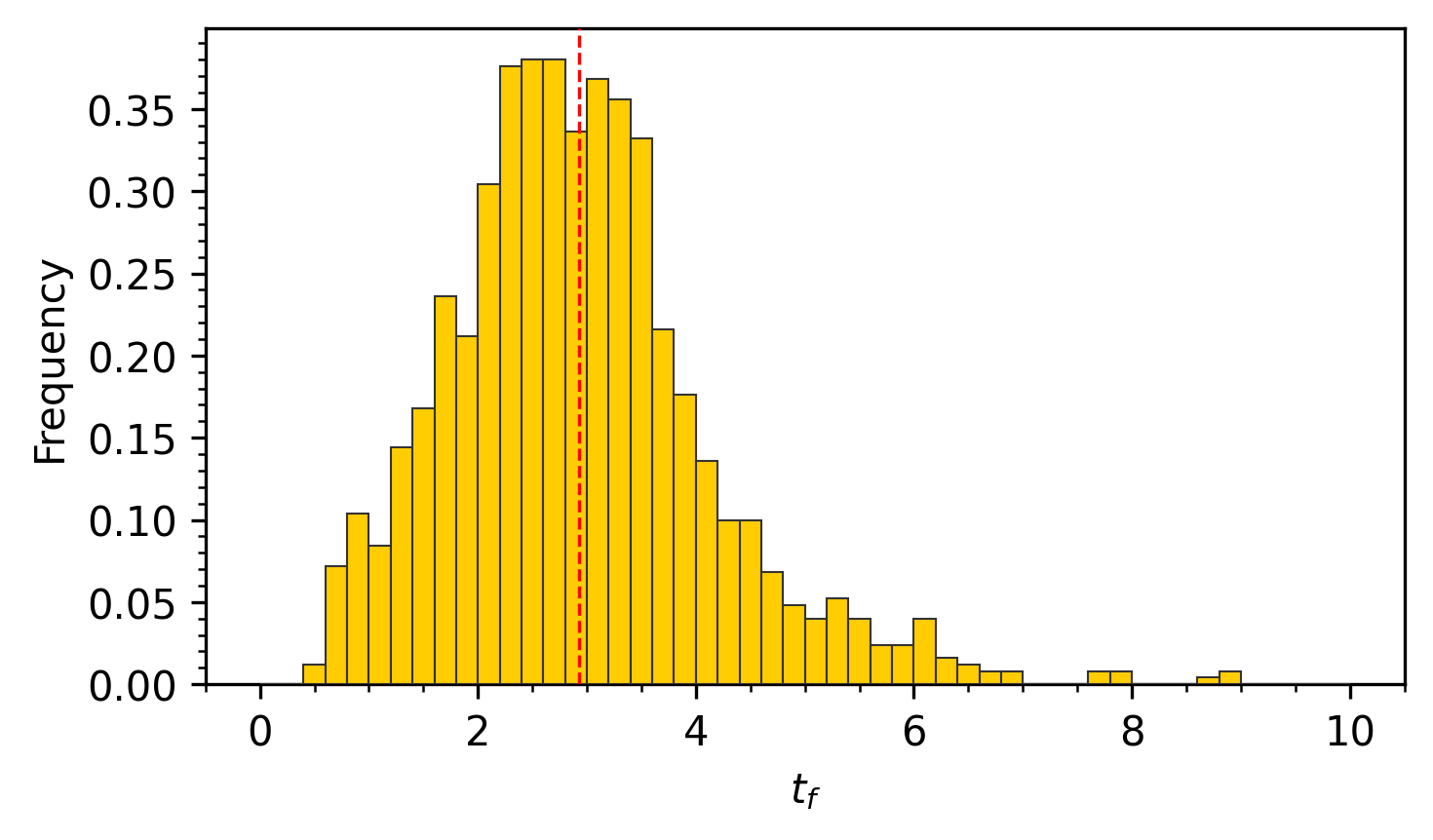}
	\caption{Histogram of forecast times $t_f$, defined by Eq.~\eqref{eqn:err}, in units of Lyapunov time for a total of $1,250$ trials using the $2,000$-node neuromorphic network. 
    An average of $2.93$ Lyapunov times is achieved (indicated by the dotted red line), with a standard deviation of $1.21$ Lyapunov times.}
	\label{fig:forecasttimes}
\end{figure}

Fig.~\ref{fig:in_out} shows the input--output mapping of a $2,000$-node, $47,946$-edge neuromorphic network used as a reservoir for dynamic RC for Lorenz63 forecasting. Input and output signals are respectively delivered to and read out from $80$ and $1,919$ randomly selected nodes.
This network and setup was used to autonomously forecast the Lorenz63 signal, with the best example shown in  
Fig.~\ref{fig:pred1}. 
The forecast time is $t_f=9.0$ Lyapunov times, after which the predicted signal deviates but still follows the overall dynamics of the Lorenz attractor.
An average forecast time of $2.9 \pm 1.2$ Lyapunov times was found over $1,250$ trials, shown in Fig.~\ref{fig:forecasttimes}, indicating the network's general performance in predicting the short-term dynamics of the Lorenz63 system. 

\begin{figure}
	\includegraphics[width=0.8\linewidth]{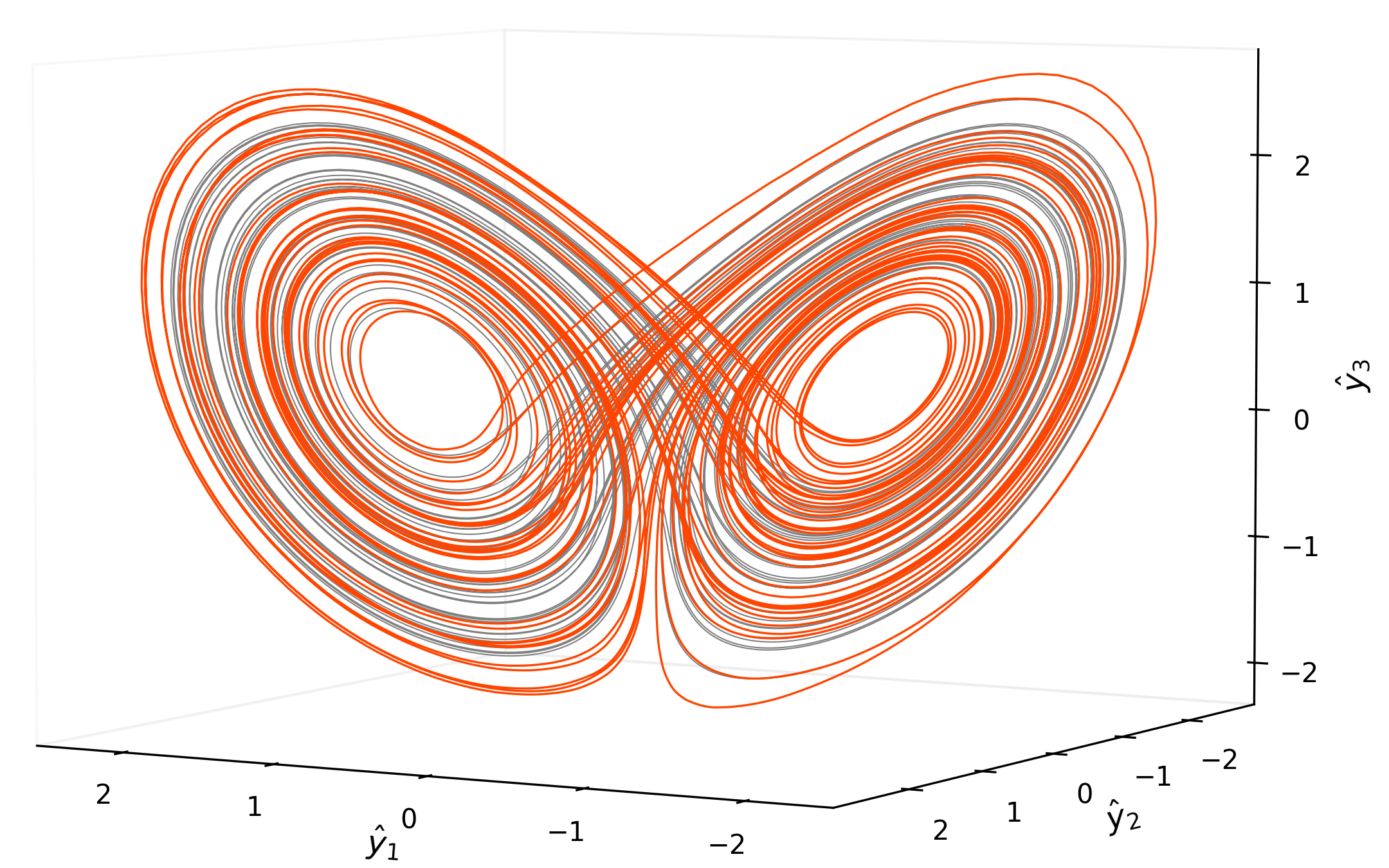}
	\caption{Predicted (orange) and true (grey) Lorenz attractor trajectory over a forecast length of $50$ Lyapunov times. 
	Data extracted from the same simulation as in Fig.~\ref{fig:in_out}. 
		}
	\label{fig:attractor}
\end{figure}

\begin{figure}
	\includegraphics[width=0.7\linewidth]{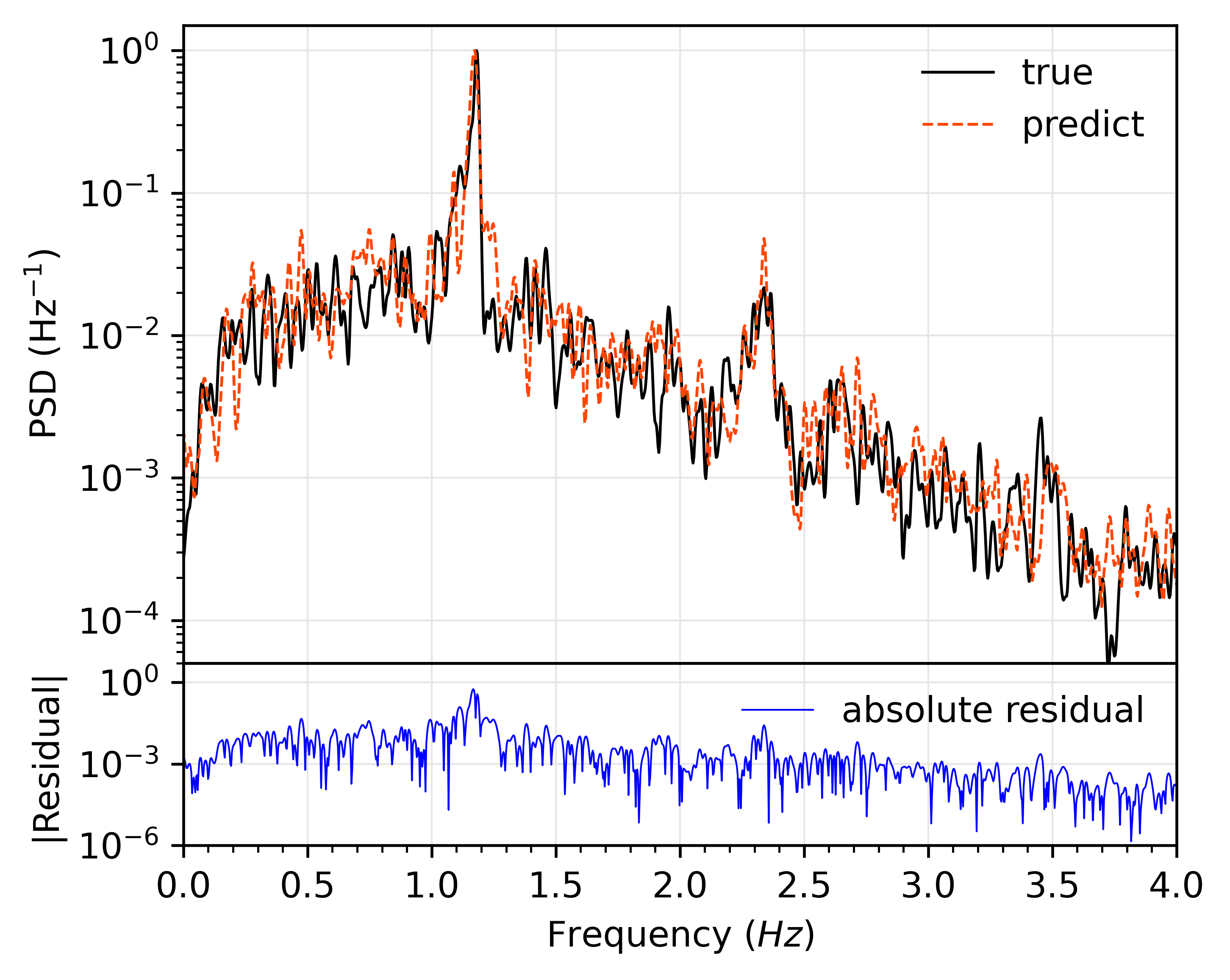}
	\caption{Normalised smoothed power spectral density (PSD) of the predicted (red) and true signal (black) of the $y_3$ component of the Lorenz system, and the corresponding absolute residual (blue), over a forecast length of $100$ Lyapunov times. 
	Data extracted from the same simulation as in Fig.~\ref{fig:pred1}, during the prediction stage, and smoothed via a Gaussian filter (with a standard deviation of $2$).}
	\label{fig:psd}
\end{figure}

With a positive Lyapunov exponent, the difference between the predicted and actual trajectory will always diverge as an unavoidable consequence of chaotic dynamics. Besides forecasting individual trajectories it is desirable to also reproduce the statistical long--term behaviour of the dynamical system. 
The simulation from Fig.~\ref{fig:pred1} was extended to $100$ Lyapunov times to produce the chaotic attractor in Fig.~\ref{fig:attractor}; over this long--term timescale, the prediction appears to qualitatively replicate the Lorenz attractor dynamics, and in particular remains stable. For a quantitative assessment of long--term forecasting, the normalised power spectral density (PSD) of the $y_3$-variable of the Lorenz system is compared in Fig.~\ref{fig:psd}, where
a high degree of concordance between the PSDs of the predicted and true Lorenz signals is evident across the majority of the frequency spectrum.

This study utilised a network of $2,000$ nodes, which is relatively large compared with other RC studies \cite{kosterDatainformed2023, akiyamaComputational2022}.
However, physical neuromorphic reservoirs typically have network sizes an order of magnitude larger, up to millions of physical nodes \cite{Stieg2011,diaz-alvarezEmergent2019,hochstetterAvalanches2021,Milano_2022,zhuMemristive2023,vahlBraininspired2024}, and they are vastly more energy efficient than their software simulator counterparts.
Due to the network size and the necessity to model each dynamical interaction within $\cal{N}$,
the simulations can be more compute intensive when compared with standard RC methods. 
However, when viewed as a simulation of a physical system, the results are valuable for optimising the design of physical neuromorphic RC systems for complex tasks such as multivariate time series forecasting. 
For example, a meta--learning scheme can be implemented for meta--parameter optimisation (e.g. $W_{\rm in}, \, \mathbf{b}_{\rm in}$) using methods such as simulated annealing \cite{zhuMemristive2023}.
Note that in real world applications, all nonlinear information processing would be completed implicitly by the physics of the neuromorphic device itself; the only computation necessary would be linear regression on the single output layer. 


\subsection{Induced Nonlinear Dynamics} \label{sec:nonlin}
\begin{figure}
	 	\includegraphics[width=\linewidth]{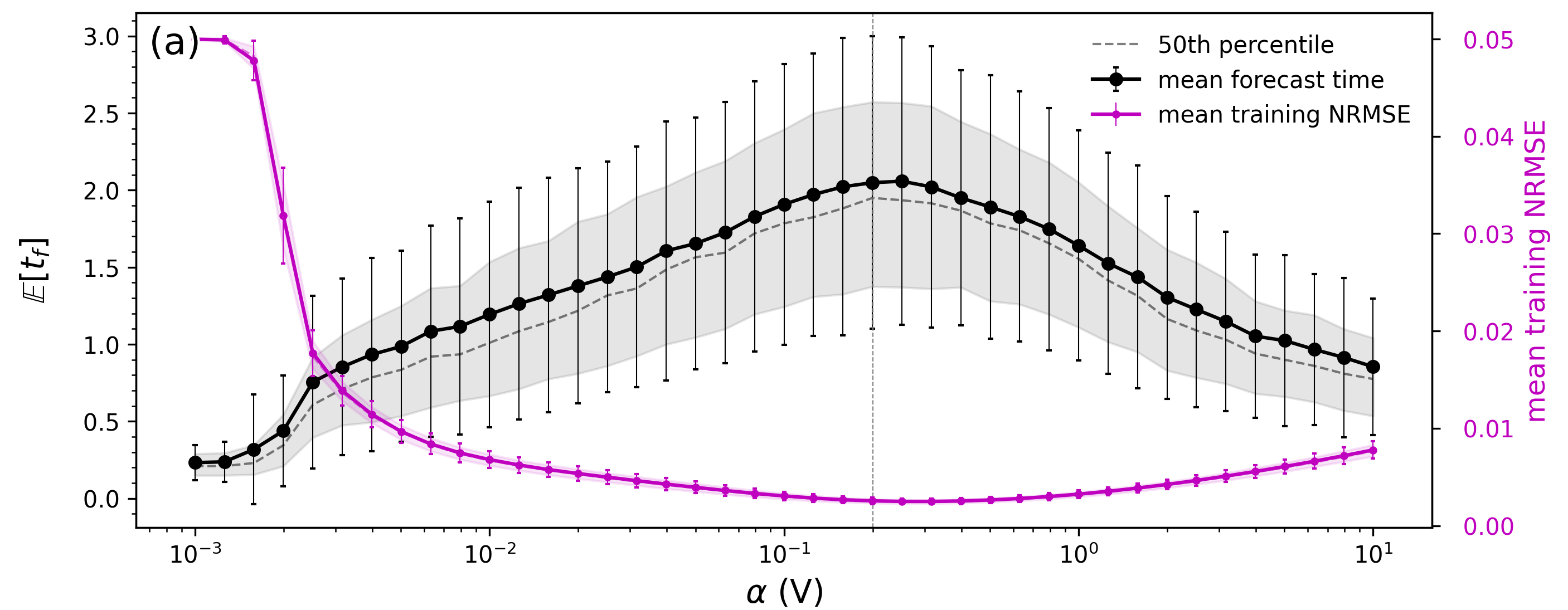}
	 	\includegraphics[width=\linewidth]{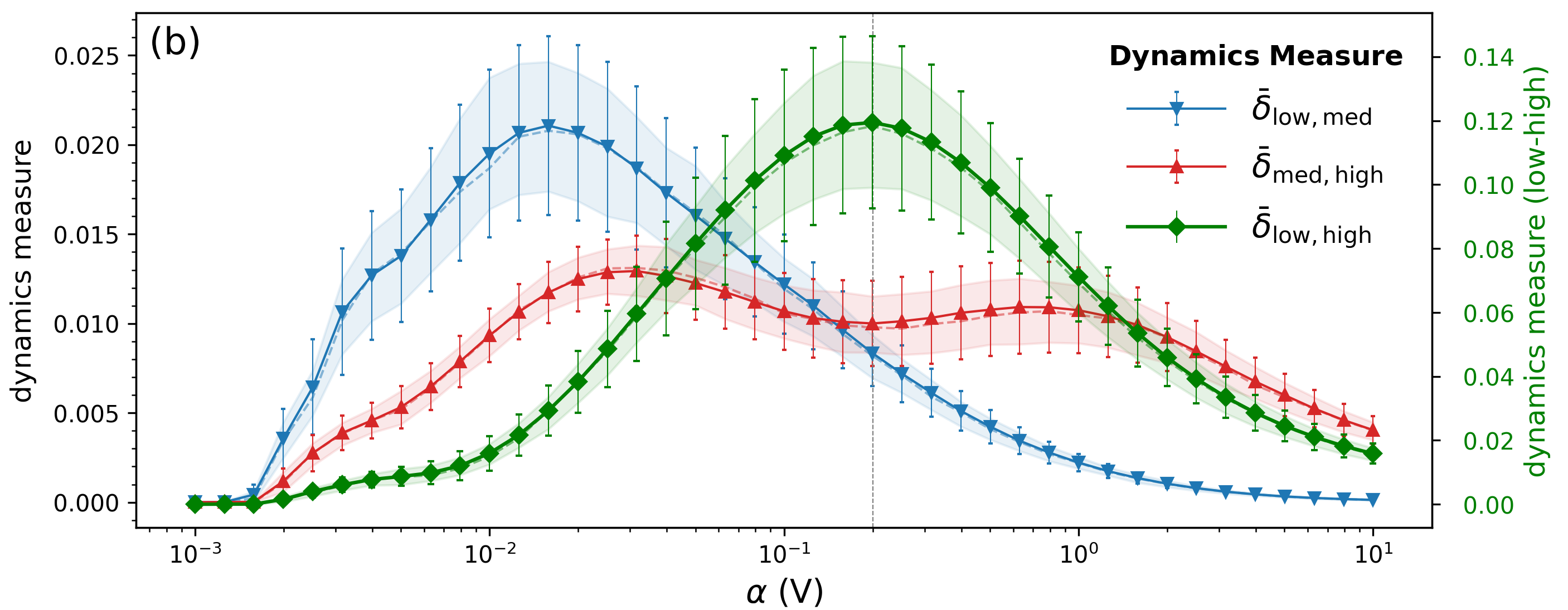}
	\caption{
	Influence of input scaling parameter $\alpha$ on forecasting performance and reservoir dynamics. (a) Mean forecast times (black) and mean training NRMSE (pink) as a function of $\alpha$; (b) corresponding dynamics measure (as defined in Eq.~\eqref{eq:avg_dynam}).
	As the low-to-high conductance state changes are one order of magnitude more frequent, the corresponding dynamics measure (green) is shown on a different scale. 
	The dotted line at $\alpha=0.2$\,V indicates the default $\alpha$ used in previous figures. 
    Simulations were performed using a neuromorphic network with $500$ nodes and $9,905$ edges. 
	Each data point in (a) was obtained from $1,500$ trials, and $100$ trials for (b).
	}
    \label{fig:nlm}
\end{figure}

\begin{figure*}
	\centering
	\includegraphics[width=0.99\linewidth]{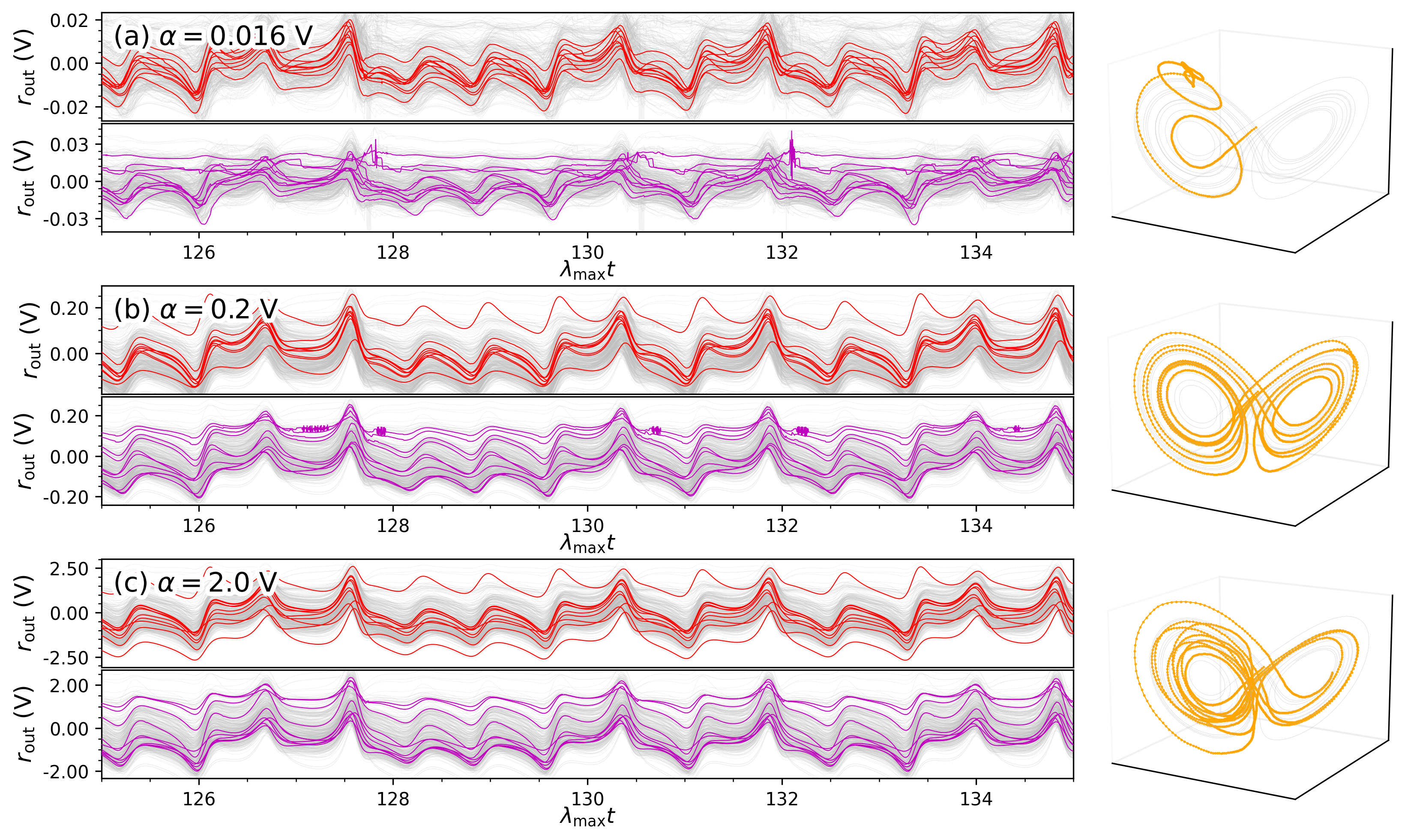}
	\caption{
		Representative behaviour of readout signals with high (red) and low (magenta) $\omega$ values ($W_{\rm out}$ weightings), and its corresponding predicted attractor (orange) with respect to input scaling parameters $\alpha$, at values of (a) $0.016\s{V}$, (b) $0.2\s{V}$, (c) $0.4\s{V}$ and (d) $2.0\s{V}$. 
		The forecast times are $t_f=1.30$, $t_f=1.64$ and $t_f=1.69$ Lyapunov times, respectively. The corresponding attractors evolve for $10$ Lyapunov times.
	}
	\label{fig:readouts_w_scale}
\end{figure*}
From Sec.~\ref{sec:dynamicRC} and Eq.~\eqref{eq:M}, it is evident that the neuromorphic network's nonlinearity and thus functionality stems from its memristive edge dynamics; hence, network performance is expected to be influenced by dynamical activity.  

From Eqs.~\eqref{eq:dx} and \eqref{eq:g}, conductance evolution is driven by the voltage difference across memristive edges. 
On a network level, this implies that nonlinear dynamics can be directly controlled (and hence optimised) by the external voltage amplitude $\alpha$ of the input signal. 
Fig.~\ref{fig:nlm}(a) shows the mean of the forecast times, $\mathbb{E}[t_f]$, as a function of $\alpha$. 
The mean forecast time exhibits a clear peak at $\alpha \approx 0.2\,$V. 
In Fig.~\ref{fig:nlm}(b) it is seen that the training error achieves its minimal Normalised Root Mean Square Error (NRMSE) at the same value of $\alpha \approx 0.2\,$V. This supports our choice for the default value $\alpha=0.2$V used in the simulations.    

As shown in Sec.~\ref{sec:dynamicRC}, nonlinearity within this system is a consequence of the memristive edge dynamics.
To quantify this, a \textit{dynamics measure} is introduced, based on changes in the conductance states of each memristive edge over a dynamical timescale of one Lyapunov time; high, medium and low conductance state changes were counted over time (cf. Fig.~\ref{fig:conductance}). 
Formally, we define the dynamics measure $\delta_{a,b}^{i,j}$ for counting the number of transitions between conductance states $X_a$ and $X_b$ across the memristive edge between node $i$ and $j$ as  
\begin{widetext}
\begin{equation}
    \delta_{a,b}^{i,j} = \frac{1}{T}\sum_\ell \left([|x_{ij}(t_\ell)|\in X_{a}][|x_{ij}(t_{\ell+1})|\in X_{b}] + [|x_{ij}(t_\ell)|\in X_{b}][|x_{ij}(t_{\ell+1})|\in X_{a}]\right), \label{eq:dynam}
\end{equation}
\end{widetext}
across all time $t_{\ell}$ during training through increments of Lyapunov times (i.e.  $t_{\ell+1}-t_{\ell}=\lambda_{\rm max}^{-1}$), normalised by the total number of training timesteps $T$.
Here $[\cdot]$ denote Iverson brackets, where $[P]$ is defined as 1 if statement $P$ is true, and 0 otherwise.  
The three conductance states are determined by their corresponding $x$ values, with intervals $X_{\rm low} = [0,0.55)$, $X_{\rm med}=[0.55,2/3]$ and $X_{\rm high} = (2/3,1]$ for the low, medium and high conductance states, respectively.
Averaging across all $K$ edges in the network, the overall network dynamic measure $\bar{\delta}_{a,b}$ is thus 
\begin{equation}
	\bar{\delta}_{a,b} = \frac{1}{K} \sum_{i,j} \delta_{a,b}^{i,j}. \label{eq:avg_dynam}
\end{equation}

The network dynamics measure is shown in Fig.~\ref{fig:nlm}(b). 
The low--med and med--high conductance changes represented by $\bar{\delta}_{\rm low,med}$ and $\bar{\delta}_{\rm med,high}$ capture dynamic activity between intermediate states, and are maximal at around $\alpha=10^{-2}\s{V}$ (with med--high state changes also persistent at $\alpha \approx 1\,$V). 
This is in fact near the onset of nonlinear dynamics, as memristive edges only evolve at a voltage greater than $V_{\rm set} = 10^{-2}\s{V}$. 
The low--high conductance changes represented by $\bar{\delta}_{\rm low, high}$ capture persistent network--scale dynamic states, corresponding to continuous formation and decay of high conductance pathways; its maximal value at $\alpha \approx 0.2\,$V also corresponds to the highest mean forecast time shown in Fig.~\ref{fig:nlm}(a).
This suggests that optimal task performance corresponds to large dynamical network activity that sweeps across many internal edge states.
Next, we consider what this means on a practical level for performing dynamic RC.

The output layer weights $W_{\rm out}$ learn nonlinear dynamical features that are most useful for predicting the Lorenz system.
Each $i$th column of $W_{\rm out}$ includes the output weighting attributed to the features produced by readout node $i$. 
Denoting the absolute sum over the $i$th column of $W_{\rm out}$ as $\omega_i$, where $\omega_i = \sum_{j=1}^{N_{\rm out}} |W_{{\rm out}}^{j,i}|$,
a high $\omega_i$ implies node $i$ produces more useful features, while low $\omega_i$ means that the $i$th feature is suppressed.
Fig.~\ref{fig:readouts_w_scale} shows examples of readout voltages during the last $10$ Lyapunov times of training for three different values of $\alpha$, with readout signals extracted from nodes $i$ with the highest ten $\omega_i$ values shown in the top panel (red), and readout signals from the smallest (near zero) ten $\omega_i$ values below (magenta), with their corresponding autonomously predicted Lorenz attractor (orange) pictured on the right. 
It is clear that all cases in Fig.~\ref{fig:readouts_w_scale} suppress readout signals with high--frequency fluctuations, suggesting that such nonlinear features are not useful for learning the Lorenz63 system. 

The nonlinear features across different $\alpha$ also show qualitative differences. 
In Fig.~\ref{fig:readouts_w_scale}(a), the value of $\alpha$ was chosen to match the onset of nonlinearity, at the maxima of the low--med dynamics measure with $\alpha=0.016$V; this is also the case with the largest amount of high--frequency fluctuations overall. 
These fluctuations are not as evident in the two cases (Fig.~\ref{fig:readouts_w_scale}(b) and (c)), with $\alpha=0.2\s{V}$ chosen to match the optimal forecasting time, and with a high value of $\alpha=2.0\s{V}$ which exhibits relatively smoother readouts.  
These are produced from the persistent dynamics in the formation and decay of conductance paths within the network, shifting the weights within the $M$ matrix gradually at every timestep.
The memristive edges evolve across all available conductance states (low through medium to high conductance state, and back), thereby maximising the use of available internal degrees of freedom.
This favourable nonlinearity is conducive for good network performance, whereas the high--frequency fluctuations represent less useful nonlinearity that is suppressed by $W_{\rm out}$.
The corresponding predicted attractor in each case shows that the most useful nonlinear dynamical features are produced away from very low and very high values of $\alpha$, where the predicted attractor becomes unstable.


\subsection{Memristor Transport Dynamics}\label{sec:transdynam}

\begin{figure*}
	\centering
	\includegraphics[width=0.95\linewidth]{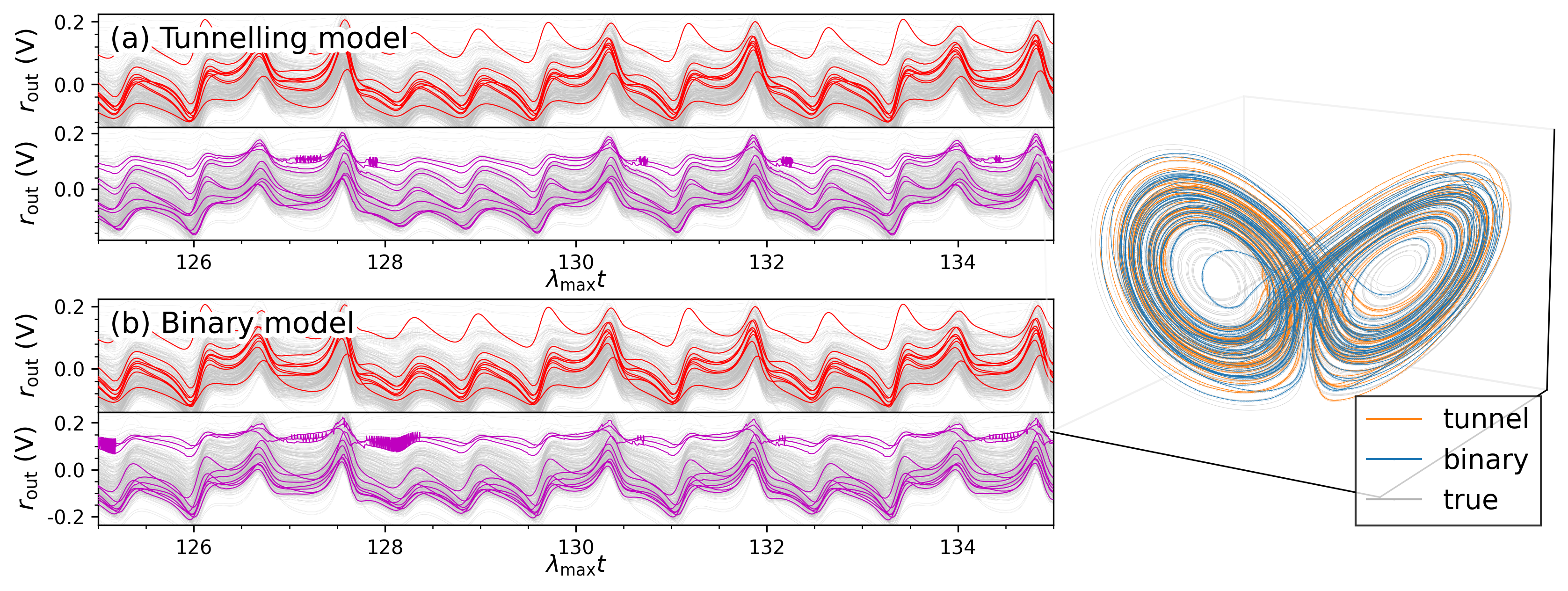}
	\caption{Representative readout signals with high (red) and low (magenta) $\omega$ weightings, for two memristive models: (a) tunnelling 
		(corresponding to the same plot for $\alpha=0.2\s{V}$ in Fig.~\ref{fig:readouts_w_scale}) and (b) binary.
		Forecast times are $t_f=1.64$ and $t_f=1.66$ Lyapunov times, respectively. The corresponding attractors (right) evolve for $50$ Lyapunov times. }
	\label{fig:readouts_w_bin}
\end{figure*}

The memristor model, Eq.~\eqref{eq:g}, utilised for this study includes electron tunnelling transport in addition to ballistic transport.
As shown in Fig.~\ref{fig:conductance}, tunnelling introduces intermediate conductance states that do not exist in the simpler binary model.
Knowing that the high--frequency fluctuations arise from low to medium conductance state changes (as previously seen in Fig.~\ref{fig:readouts_w_scale}(a)), it is reasonable to question whether they are caused by tunnelling, that is, if the neuromorphic network's capabilities are limited by the inherent physics of the system.

Fig.~\ref{fig:readouts_w_bin} reveals, however, that the high--frequency fluctuations are even more pronounced in the binary model. 
The shown examples of both models (simulated with the same parameters) share similar below--average forecast times of $t_f\approx 1.6$ Lyapunov times, and their predicted attractors exhibit similar long--term dynamics (up to to 50 Lyapunov times); both successfully reproduce the Lorenz dynamics, only deviating in regions with relatively little training data (at the edges of the attractor).
Thus, the high--frequency nonlinear features do not originate from tunnelling dynamics \textit{per se}, but rather appear to arise as a result of rapid switching of conductance states,
fluctuating around the voltage threshold $V_{\rm set}$ for the onset of memristive dynamics.
Moreover, the observation that the rapid fluctuations are more prominent in the binary model suggests that the tunnelling model is able to naturally suppress these features, and this may be attributed to the many more degrees of freedom afforded by the intermediate states of tunnelling dynamics (see Fig.~\ref{fig:induced_flux_binary} in Appendix).
Nevertheless, both models still yield similar average short--term prediction accuracy (see Fig.~\ref{fig:bin_hist} in Appendix) due to the suppression of these features via ridge regression in the output layer.
See Appendix \ref{app:edge} for further details on how these high--frequency fluctuations arise from memristive edges in both models.

\begin{figure}
	\centering
	\includegraphics[width=0.9\linewidth]{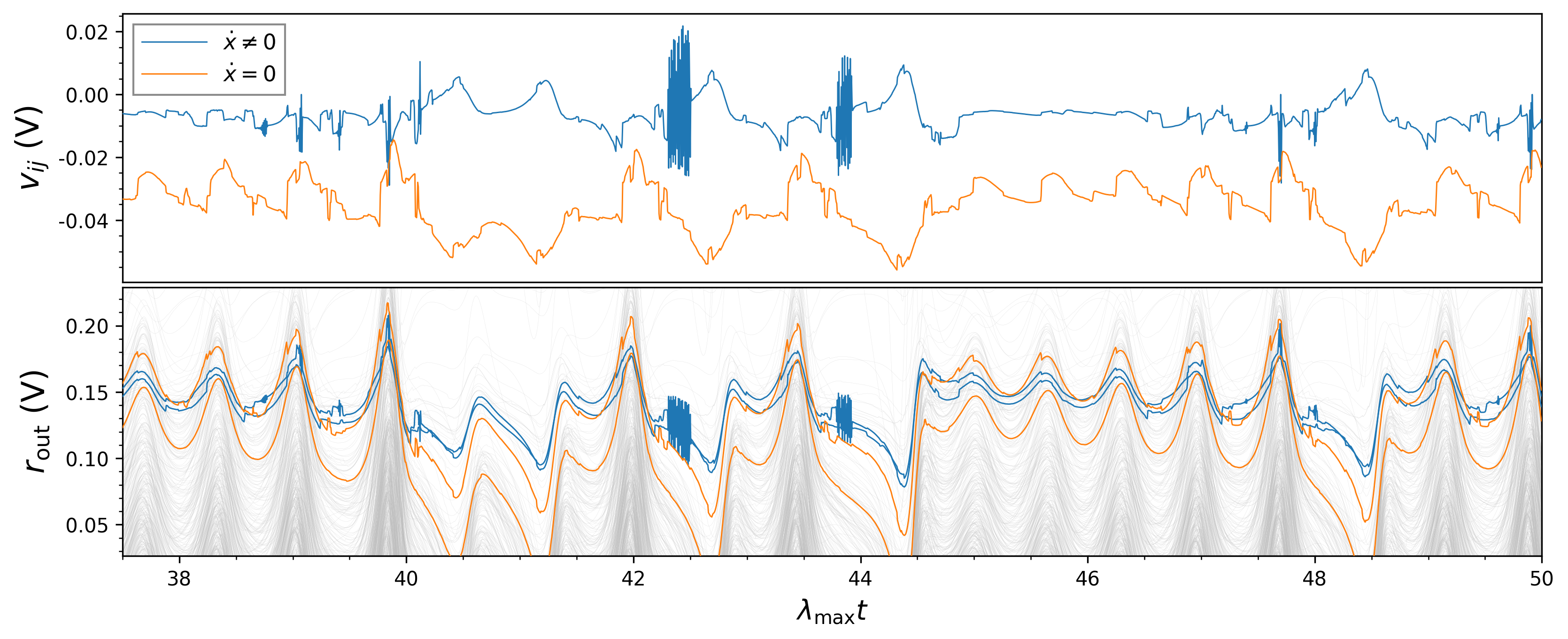}
	\caption{Examples of (top) memristive edge voltages with and without dynamic activity ($\dot x \neq 0$ and $\dot x =0$, respectively) and (bottom) their corresponding node voltage readouts, showing that high--frequency nonlinear features may be present even in nodes connected to edges that are not evolving (orange), as compared to a typical nonlinear voltage signal (blue). 
	All simulation parameters are the same as in Fig.~\ref{fig:readouts_w_scale} with $\alpha=0.2\s{V}$.}
	\label{fig:no_flux}
\end{figure}

The intermediate conductance state changes (low--med, med--high) do not alone explain all fluctuating dynamics in the network; by studying dynamics at the individual memristive edge level, these high--frequency fluctuations can also be observed in nodes connected to edges which have no dynamical activity.
This is shown in Fig.~\ref{fig:no_flux}, where high--frequency fluctuations can also be observed in edges without any dynamics (i.e. $\dot{x}_{ij}=0$ at edge between node $i$ and node $j$). 
For a more detailed analysis, see Appendix \ref{app:edge}.
In summary, these high--frequency dynamical features are caused in part by the network connectivity distributing nonlinear dynamics amongst all connected parts, effectively acting as a multiplier of nonlinearity, hence allowing network nodes to exhibit nonlinear dynamics even when corresponding edges have no internal dynamical activity themselves.

Overall, nonlinear dynamical features are induced by the complex interactions between memristive edge evolution and network connectivity, resulting in the formation and destruction of conductance paths and feedback loops which cannot be easily inferred unless fully numerically simulated (since a closed--form solution does not exist). 
From these results, it is expected that proximity of non--active network components to dynamically active components should affect how dynamical features are generated and distributed throughout the network; this is explored next.


\subsection{Node Interactions}\label{sec:node_inter}

\begin{figure*}
	\centering
	\includegraphics[width=\linewidth]{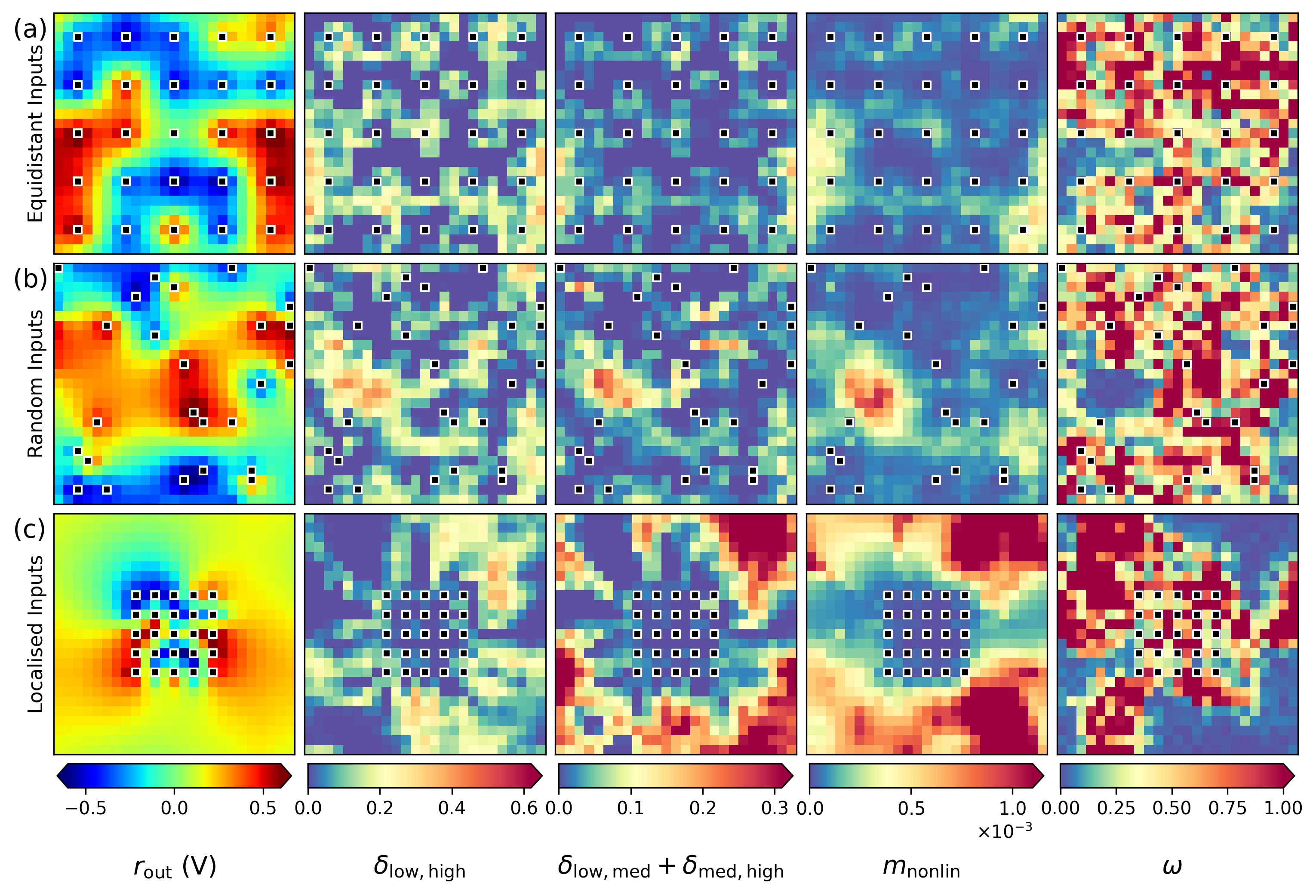}
	\caption{
	Heat map of selected time-averaged measures on a 2D $25\times 25$ doubly periodic lattice network for different input electrode node placements: (a) uniformly equidistant inputs, (b) randomly selected inputs, and (c) localised inputs. 
	The first column shows the readout voltages $r_{\rm out}$ averaged over time.
	The second column shows the dynamics measures $\delta_{\rm low,high}$ for conductance state transitions between low and high conductance states,
	the third column shows dynamics measures $\delta_{\rm low,med}+\delta_{\rm med,high}$ for low-med and mid-high conductance states changes (defined in Eq.~\eqref{eq:dynam}),
	the fourth column shows the nonlinearity measure $m_{\rm{nonlin}}$ (defined in Eq.~\eqref{eq:m_nonlin}),  
	and the fifth column shows the corresponding $\omega$ weights. 
    Indices omitted in all labels for clarity. 
	}\label{fig:lattice}
\end{figure*}
\begin{figure*}
	\centering
	\includegraphics[width=0.9\linewidth]{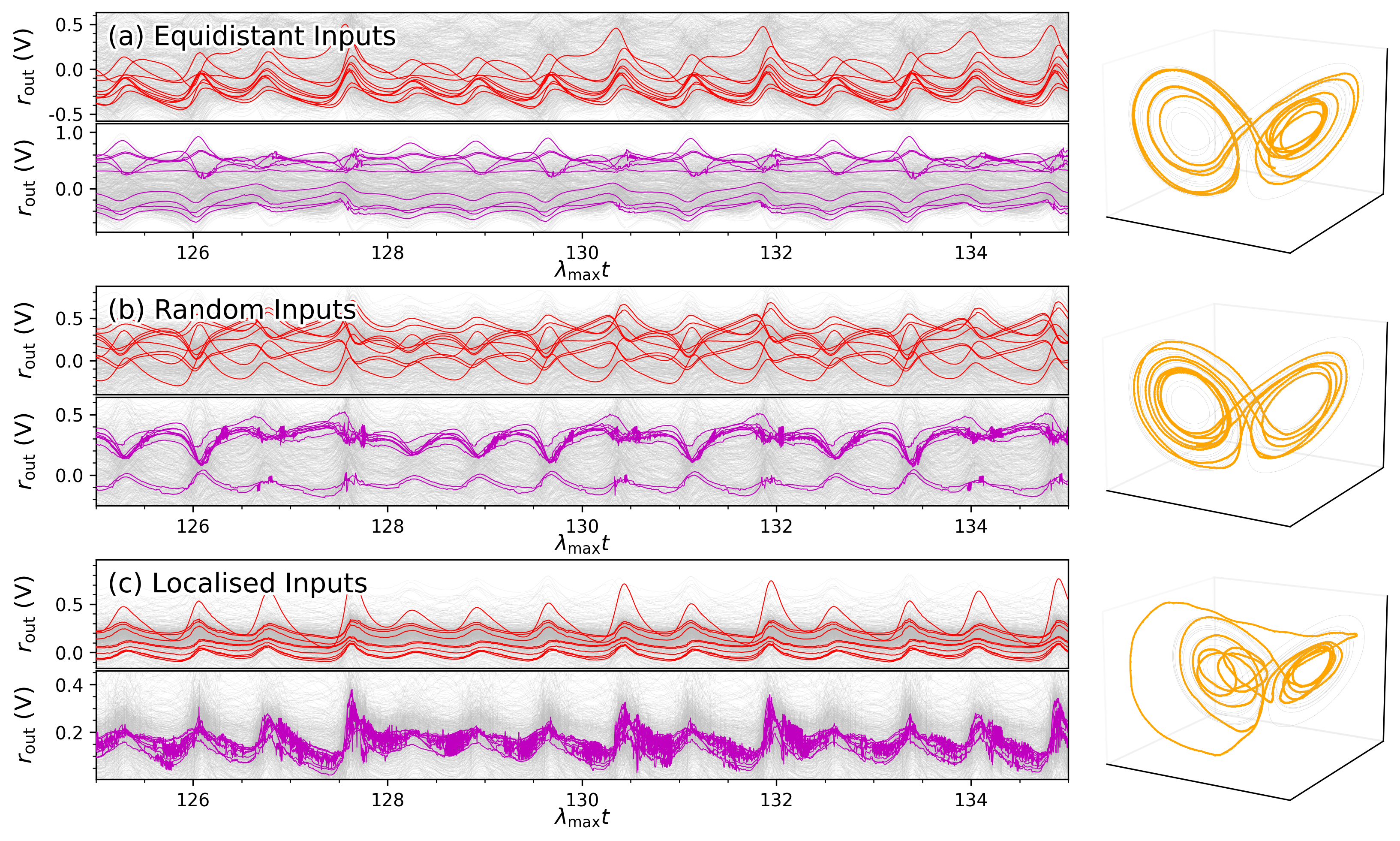}
	\caption{
		Representative readout signals with high (red) and low (magenta) $\omega$ weightings, and their corresponding attractors (orange), for different input node placements, for the experiments on the doubly periodic lattice shown in Fig.~\ref{fig:lattice}. (a) uniformly spaced inputs, (b) randomly selected inputs and (c) localised inputs. 
		The respective forecast times are $t_f=3.41$, $t_f=1.68$ and $t_f=1.58$ Lyapunov times. The corresponding attractors evolve for $10$ Lyapunov times.
		}
	\label{fig:readouts_w}
\end{figure*}
Local node interactions are influenced by the input electrode node placements.
This section investigates how proximity of readouts to input electrodes affects nonlinear dynamics, and how it may be possible to infer input node allocations that promote good forecasting performance.

To aid analysis and visualisation of node interactions,
a homogeneous square 2D lattice network was constructed. 
This network consists of $25 \times 25 = 625$ nodes, with each node connected to $4$ other nodes (implying $1,250$ edges), with doubly periodic boundary conditions. 

We introduce a \textit{nonlinearity measure}, $m_{\rm nonlin}^{i,j}$, 
defined as the mean squared difference between the node readouts $r_{\rm out}^{i,j}$ of the nonlinear model and the readouts $r_{\rm out}^{L;i,j}$ of the closest linear model, normalised by the mean squared readout signals, 
\begin{equation}
	m_{\rm nonlin}^{i,j} = \frac{\left\langle {\left|{r}_{\rm out}^{i,j} - {r}_{\rm out}^{L;i,j}\right|^2} \right\rangle}{
		\left\langle {||\mathbf{r}_{\rm out}||^2} \right\rangle
	}.
    \label{eq:m_nonlin}
\end{equation}
On the overall network level, this nonlinearity measure $\bar{m}_{\rm nonlin}$ would be 
\begin{equation}
	\bar{m}_{\rm nonlin} = \frac{\left\langle {||\mathbf{r}_{\rm out} - \mathbf{r}_{\rm out}^{L}||^2} \right\rangle}{
		\left\langle {||\mathbf{r}_{\rm out}||^2} \right\rangle
	}.\label{eq:avg_m_nonlin}
\end{equation}
The readout $\mathbf{r}_{\rm out}^L$ is obtained from a linear model for which the internal state variables are static with $\dot{\mathbf{x}}=0$ for all memristive junctions, effectively rendering them linear resistors. 
Consequentially for very small values of $\alpha<0.001\s{V}$, $\bar{m}_{\rm nonlin} =0$ as voltage signals is too small to form any conductance paths (i.e. $\dot{\mathbf{x}}=0$).
Likewise for large values of $\alpha>10\s{V}$, the corresponding $\bar{m}_{\rm nonlin}$ are also approximately zero, as all memristive edges are oversaturated with voltage, effectively also resulting in a near linear network with $\dot{\mathbf{x}}\approx0$ (see Fig.~\ref{fig:nonlin_meas} in Appendix). 

Fig.~\ref{fig:lattice} shows heat maps of the time--averaged readout voltages (column 1), along with the low--high dynamics measure (column 2), the sum of the low--med and med--high dynamics measures (column 3), the nonlinearity measure defined in Eq.~\eqref{eq:m_nonlin} (column 4), and weight matrix $\omega$ values (column 5), with each pixel value representing a node that respects its relative location on the lattice network. 
Fig.~\ref{fig:lattice}(a) uses uniformly distributed, equidistant input electrode nodes, while Fig.~\ref{fig:lattice}(b) and (c) use randomly placed and localised input nodes, respectively.
Comparison of these heat maps suggests that the nonlinearity measure is largely correlated with the dynamics measures (especially the low--med and med--high dynamics), and that nonlinearity is anti--correlated with $\omega_i$.
This anti--correlation implies suppression of highly nonlinear and certain highly dynamical readout nodes by the output layer, which means on a local level some highly nonlinear regions are not useful for the prediction task.

Fig.~\ref{fig:lattice} reveals that nonlinear regions appear mostly in regions which are far away from input electrode nodes.
This is more obvious in Fig.~\ref{fig:lattice}(c), where by clustering all input nodes together in one region, it becomes clear that the nonlinear parts are the areas away from input nodes.
The anti--correlation between $\omega_i$ and dynamics measures is also clear in this example, where the highest $\omega_i$ weights are mostly in areas near the input nodes.
Every input electrode node has its own area of influence, and although this can be affected by other input nodes, on average the further away a readout node is to this input, the less influence this input has on the readout.
These far--away nodes dynamically behave in a manner increasingly more different from those of the input electrode nodes.

Note, however, that this anti--correlation is not exact, nor is the correlation between the nonlinearity and dynamics.
This is due to the aforementioned network connectivity effects; it is most obvious in the localised input example, where there exists regions with relatively little dynamics but high nonlinearity, and these regions do appear smooth and connected, tending to distribute amongst close neighbouring network components.

The suppressed highly nonlinear nodes also exhibit high--frequency fluctuations in their readout signals, which is the unfavourable nonlinearity encountered previously in Fig.~\ref{fig:readouts_w_scale}.
Fig.~\ref{fig:readouts_w} shows the corresponding readout signals of the above three examples in Fig.~\ref{fig:lattice}, with the readout signals corresponding with the largest (red) and smallest (magenta) ten $\omega_i$ weights both emphasised. 
As seen previously in Fig.~\ref{fig:readouts_w_scale}, high--frequency fluctuations can be observed in Fig.~\ref{fig:readouts_w} in signals with small $\omega_i$ values.
As expected, these suppressed nonlinear features are also much more prominent in Fig.~\ref{fig:readouts_w}(c) with the clustered inputs.
This implies the unwanted nonlinear readouts arise from nodes distant from the inputs, restricting them to only evolve between the low and medium conductance states as indicated by the low--med dynamic measure, which correlates with the appearance of nonlinear signal that exhibit high--frequency fluctuations.

Due to lower presence of unwanted nonlinear features, the examples with random and equidistant inputs both perform better than the clustered input example, as seen from the Lorenz attractor predictions in Fig.~\ref{fig:readouts_w}. 
This impact of input node placement remains true for the heterogeneous neuromorphic network (see Fig.~\ref{fig:cluster_hist} in Appendix).
Note that with a high network density, randomly selecting nodes leads to an input electrode node placement in which nodes are on average separated by the same number of edges. 
Even in the sparse lattice network structure there are only minor differences between equidistant and the randomly selected input nodes (cf. Fig.~\ref{fig:lattice}(a) and Fig.~\ref{fig:lattice}(b)). 

To maximise the area of influence of each readout node over time, and hence minimising the occurrence of less useful nonlinear features, the input voltage should remain sufficiently large on average. 
This can be achieved with the input bias voltages. 
Instead of sampling values in $\mathbf{b}_{\rm in}$ uniformly from $[-b,b]$, sampling from $[-b, -b/2]\cup [b/2, b]$ ensures that no input node will have low voltages. 
On average this does have a noticeable effect in the Lorenz prediction task, where for the $500$-node network an average forecast time of $2.12$ Lyapunov times can be observed with the new bias sampling, as opposed to $1.95$ when the bias is sampled form the whole range $[-b,b]$ (cf. Fig.~\ref{fig:ablation} in Appendix). 

In summary, proper input electrode node number and placement allows maximal usage of the given network, by ensuring all readouts are within the area of influence of the inputs such that less useful nonlinear features are minimised. 
Practically, in a sufficiently dense network, random allocation of input electrode nodes is very similar to evenly distributed input nodes. 
This motivates our choice to choose the number of input nodes as around $5$\% of all available nodes --- with roughly evenly spaced input nodes, $5$\% is sufficient to activate and continuously sustain network dynamics.

\begin{figure}
    \centering
    \includegraphics[width=\linewidth]{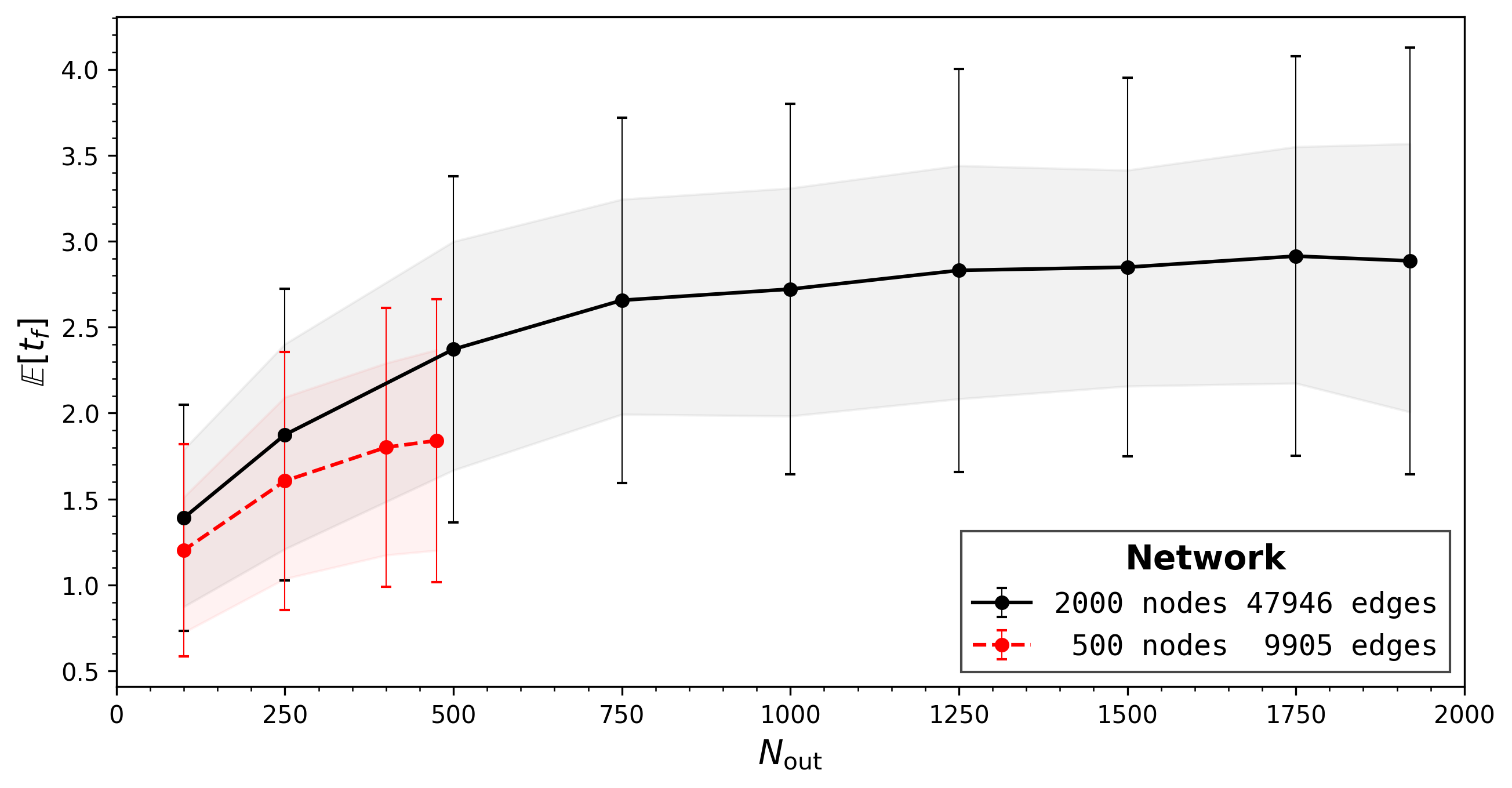}
    \caption{
    \rva{Average forecast time with respect to the number of readout electrode nodes $N_{\rm out}$, for two different sized neuromorphic networks. Shaded region indicates the 25th and 75th percentile, and error bars indicate 1 standard deviation. All data points obtained from 500 trials each.} 
    }
    \label{fig:limited_readouts}
\end{figure}

\rva{
All numerical simulations performed in this study use $\simeq 90$\% of all available nodes as readout nodes, which may limit its hardware implementation.
However, as shown in Fig.~\ref{fig:limited_readouts}, fewer readouts can still produce good forecast times;
for a $2,000$-node network, forecasting saturates above $N_{\rm out} \simeq 1,250$, corresponding to $\simeq 63$\%.
The readout nodes in Fig.~\ref{fig:limited_readouts} are selected by the size of their $\omega$ weightings (i.e. a portion of the lowest $\omega$ weighted nodes are pruned for small $N_{\rm out}$). 
Most importantly, Fig.~\ref{fig:limited_readouts} shows that for neuromorphic networks of two different sizes but with the same number of readout nodes, the larger network always outperforms the smaller network: 
with 500 readouts, the average forecast time is $\simeq 2.4$ Lyapunov times for the larger $2,000$ node network, whereas it cannot exceed 1.8 Lyapunov times for the smaller $500$-node network. 
Considering the millions of physical nodes and memristive edges in neuromorphic nanowire networks~\cite{sillin2013theoretical,diaz-alvarezEmergent2019}, the hardware implementation is expected to vastly outperform the simulations in this study for predicting the Lorenz system.
Moreover, experimental~\cite{zhu2023online} and simulation studies~\cite{Daniels_etal_2022,Monaghan_etal_2025} indicate that devices can still perform well on physical RC tasks even with a limited number of electrodes. This may be attributed in part to signal integration, since individual electrodes are contacted to many nanowires.
Further studies are warranted to determine optimal electrode number and configuration
such that they maximise the usage of neuromorphic network dynamics for physical RC.
}

\section{Conclusions}
\label{sec:conclusions}

These results validate the hypothesis that the internal and network dynamics of memristive networks can be used in a reservoir computing approach to learn complex chaotic dynamics.
It was found that the network dynamics can be optimised for predicting the Lorenz63 attractor by appropriate scaling of the external input voltage to values that are not too low or too high.
Optimal intermediate values of input voltage scaling were found to drive internal memristive states that traversed the entire dynamical range, from low to high conductance states, thereby maximising use of available internal degrees of freedom.
Interestingly, this study also revealed that some nonlinear dynamical features, such as high--frequency fluctuations, were effectively filtered out by ridge regression in the external output layer. 
This suggests that physical reservoir computing may be robust against noise and variations due to the specific internal dynamics of the memristive material system. 
On the other hand, our results also indicate that it is possible to suppress such features by ensuring electrodes provide sufficient coverage of network nodes to sustain dynamic activity across the memristive edges.

Overall, this study revealed how the nonlinear dynamics of memristive neuromorphic networks can be controlled and optimised for reservoir computing solely via adjusting the external input layer parameters.
This is promising for practical implementations of physical reservoir computing with memristive networks, which have orders of magnitude more physical nodes and memristive edges \cite{diaz-alvarezEmergent2019,sillin2013theoretical}.
Ongoing work is also exploring the optimal network structure\cite{ijcnn_unpublished}, output layer and computational frameworks beyond standard reservoir computing, to be eventually applied to real world applications on noisy, unstructured data, thereby unlocking the full potential of neuromorphic networks for modelling dynamical systems via the dynamics of the physical network.

\begin{acknowledgments}
The authors acknowledge use of the Artemis high performance computing resource at the Sydney Informatics Hub, a Core Research Facility of the University of Sydney. 
Y.X. is supported by an Australian Government Research Training Program (RTP) Scholarship. G.A.G. acknowledges support from the Australian Research Council under Grant No. DP220100931.
\end{acknowledgments}

\section*{Data Availability}
The data that support the findings of this study may be generated with the code available at \url{https://github.com/xvyh/dynamicRC_NWN}.

\appendix

\section{Parameters for the Memristor Model} \label{app:sim}
The parameters $\zeta_0$ and $\zeta_1$ in Eq.~\eqref{eq:simmons} are given by
\begin{align}
	\zeta_0 &= A \frac{3(2m_*)^{1/2}e^{5/2}(\phi/e)^{1/2}}{2h^2}, \\
    \zeta_1 &= - \frac{4\pi (2m_* e)^{1/2}}{h} \left( \frac{\phi}{e}\right)^{1/2},
\end{align}
and are comprised entirely of physical parameters of the nanowires as described in Ref.~\cite{hochstetterAvalanches2021}.
The largest and smallest resistances of the memristors are respectively assigned the values of $R_{\rm off} = 12.9\times 10^6 \s{\Omega}$, and $R_{\rm on} = 12.9\times 10^3 \s{\Omega}$.
The memristor voltage needed to start the evolution of $x$ is $V_{\rm set}=0.01 \s{V}$, and the voltage to reset the memristor state is $V_{\rm reset}=0.005 \s{V}$.
The largest flux is $\xi_{\rm max}=0.015 \s{Vs}$, the normalised critical flux value 
is $x_{\rm crit} =2/3$. 
The thickness of the electrically insulating, but ionically conducting polymer layer between nanowires is $s_{\rm max}=5\s{nm}$.
The boost parameter in Eq.~\eqref{eq:dx} is $q=10$, and the global $\dot{x}$ scaling parameter $\eta=\eta_0/\xi_{\rm max}$ is determined with $\eta_0=10$.

\section{Nonlinear Responses in Memristive Edges} \label{app:edge}
\begin{figure*}
	\centering
	\includegraphics[width=0.45\linewidth]{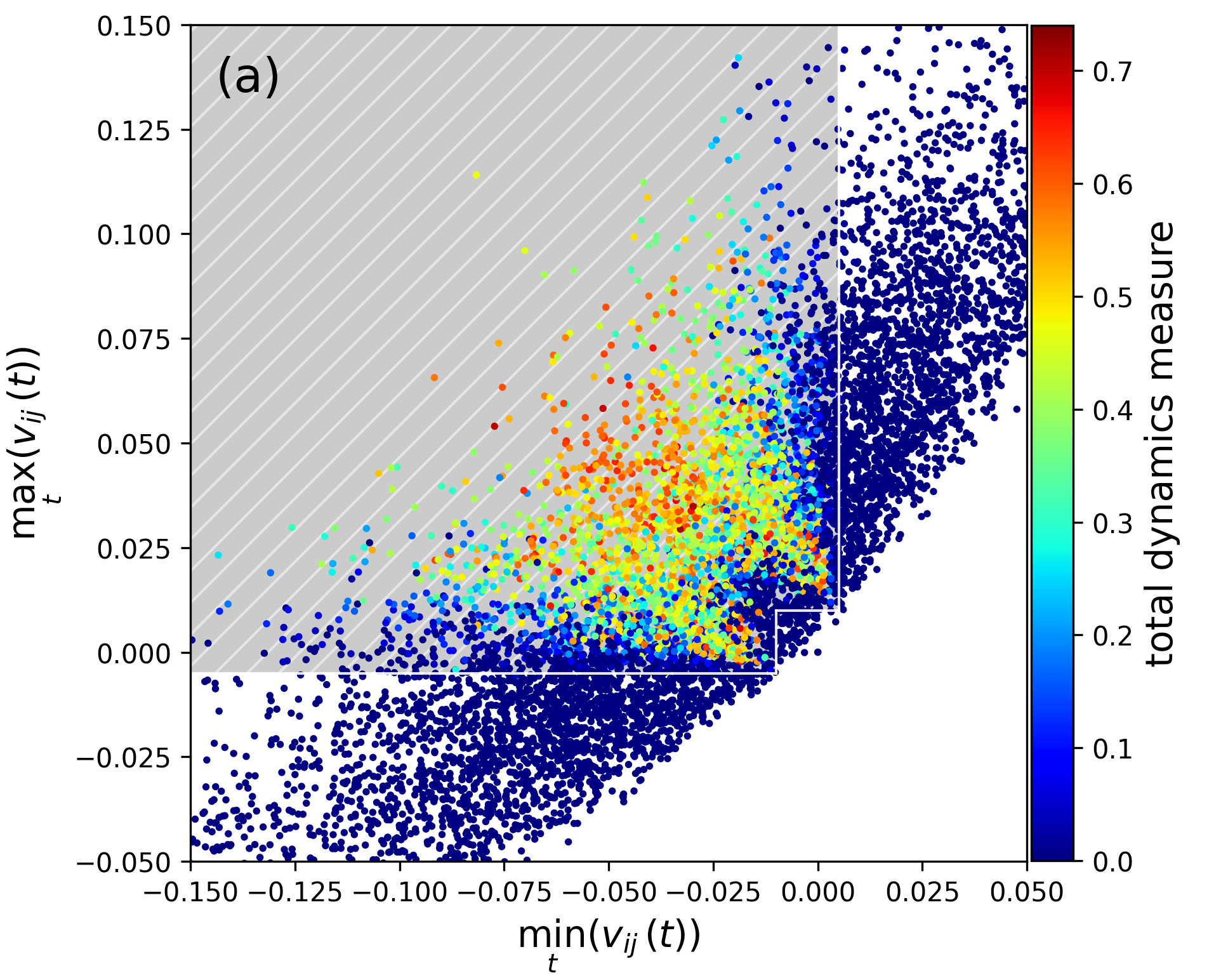}
	\includegraphics[width=0.45\linewidth]{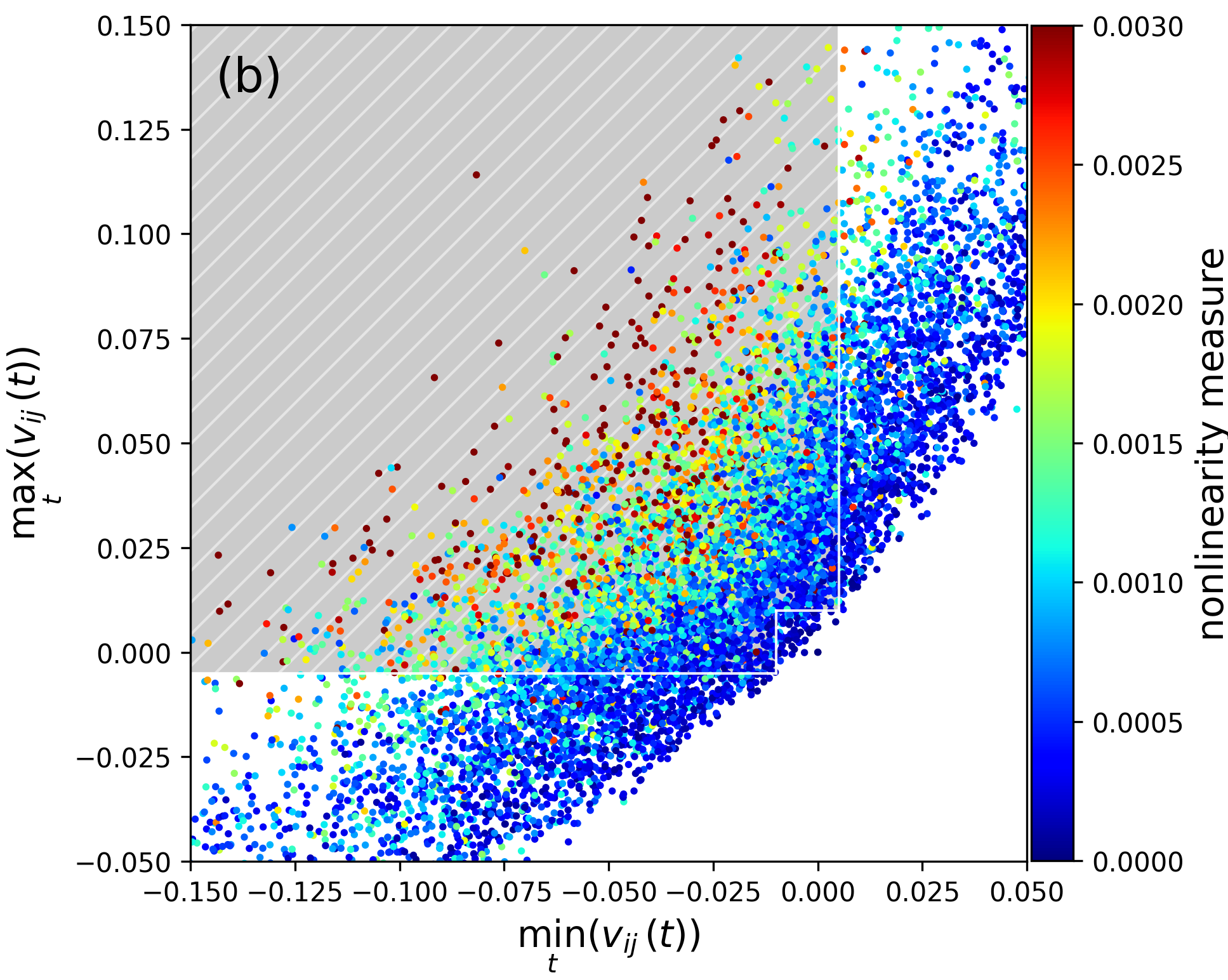}
	\caption{(a) Total dynamics measure (i.e. $\delta_{\rm low,med}^{i,j} + \delta_{\rm med,high}^{i,j}+\delta_{\rm low,high}^{i,j}$) and (b) nonlinearity measure $m_{\rm nonlin}^{i,j}$ with respect to the maximum and minimum of each corresponding memristive edge voltage signal across time. 
		The shaded regions are where edge voltage signal amplitudes can cross $V_{\rm set}$ and $V_{\rm reset}$, i.e. where $\dot{x}_{ij}$ may be nonzero. 
		All data points are obtained from the same experiment for $\alpha=0.2 \s{V}$ in Fig.~\ref{fig:readouts_w_scale}. 
		Edges connected to the input nodes are omitted, and the figures are zoomed in for clarity. 
	}
	\label{fig:jnv_scatter}
\end{figure*}

\begin{figure}
	\centering
	\includegraphics[width=\linewidth]{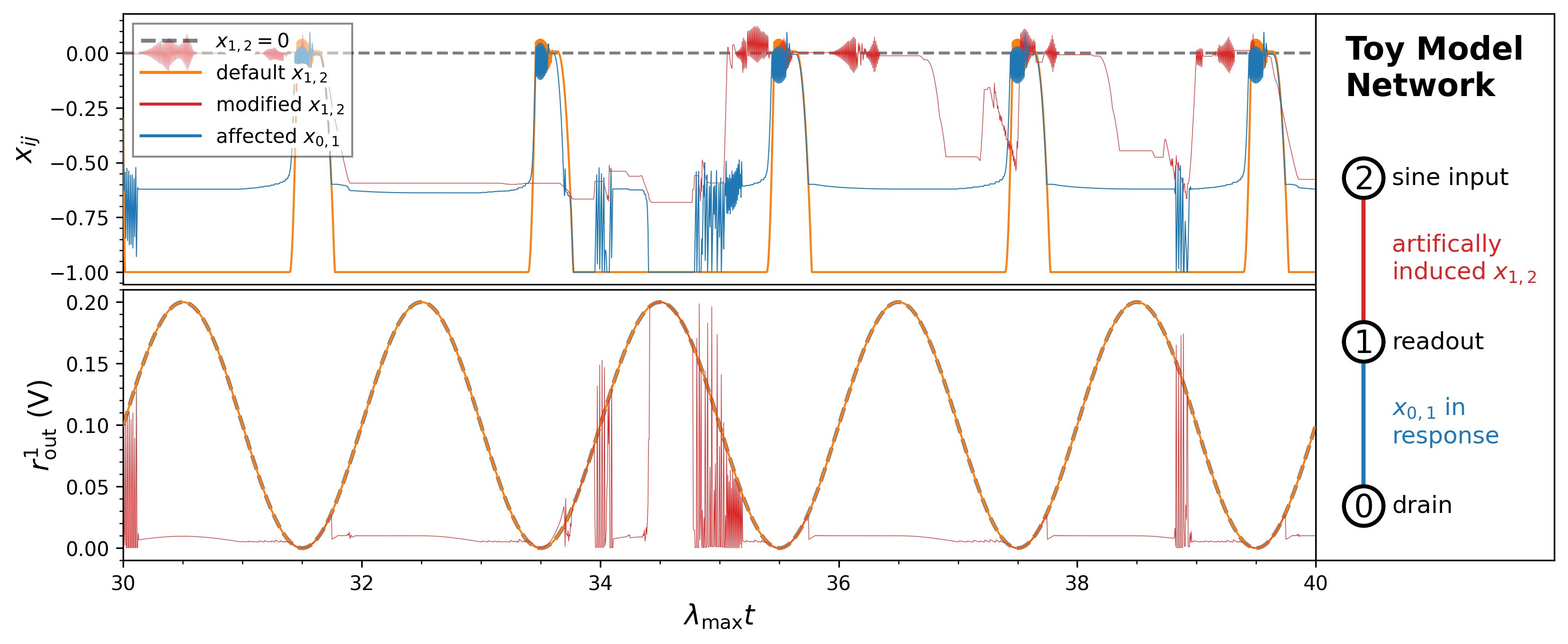}
	\caption{
		By artificially inducing a highly nonlinear (top, red) $x$ signal into the edge between node 2 and 1 of this simple 3-node toy model, unwanted nonlinear features (bottom, red) can be observed in the readouts of node 1, which are vastly different from the original (orange) signal. 
		This also affect the $x$ (top, blue) signals of the edge between node 1 and 0.
		These memristive edges uses the tunnelling model.}
	\label{fig:induced_flux_tunnel}
\end{figure}

\begin{figure}
	\centering
	\includegraphics[width=\linewidth]{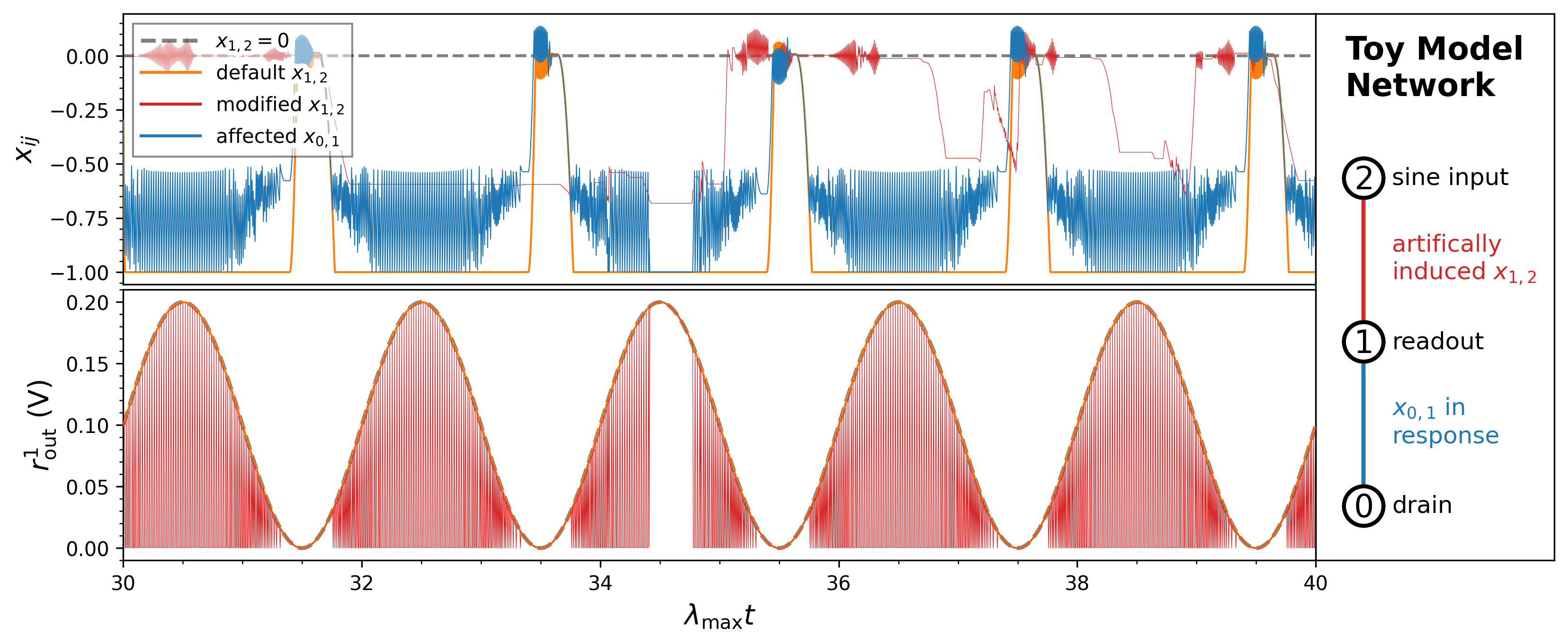}
	\caption{
		By artificially inducing a highly nonlinear (top, red) $x$ signal into the edge between node 2 and 1 of this simple 3-node toy model, unwanted nonlinear features (bottom, red) can be observed in the readouts of node 1, which are vastly different from the original (orange) signal. 
		This also affect the $x$ (top, blue) signals of the edge between node 1 and 0.
		Unlike Fig.~\ref{fig:induced_flux_tunnel}, these memristive edges uses the binary model.}
	\label{fig:induced_flux_binary}
\end{figure}
From Eq.~\eqref{eq:dx} it is evident that $x$ will only increase in amplitude when the voltage amplitude at the corresponding memristive edge crosses $V_{\rm set}$, and decrease in amplitude when it crosses $V_{\rm reset}$.
This will occur in two scenarios: if the maximum edge voltage signal (across time)  of the edge between node $i$ and node $j$ satisfies $\max_t (v_{ij}(t))>V_{\rm set}$, while the minimum edge voltage satisfies $\min_t (v_{ij}(t))<V_{\rm reset}$, or 
$\max_t (v_{ij}(t))>-V_{\rm reset}$ and $\min_t (v_{ij}(t))<-V_{\rm set}$. 
This results in two overlapping regions where dynamical activity can be observed. 
Indeed as seen in Fig.~\ref{fig:jnv_scatter}(a), the dynamic measure is only nonzero within this specified (shaded) region, and it is clear that larger ranges (i.e. larger $\max_t (v_{ij}(t))$ and smaller $\min_t (v_{ij}(t))$) of edge voltages which centre near $0$ tend to produce higher dynamics, while those near the boundary of dynamic activity (i.e. boundary of the shaded region) produce less.

However note that in Fig.~\ref{fig:jnv_scatter}(b), although nonlinearity is maximal and most frequent within the specified region, there exists numerous highly nonlinear edges outside this region. 
The fact that nonlinearity can exist in edges which do not dynamically evolve implies the network connectivity has a significant effect on the overall nonlinear dynamics, and memristive edge dynamics is only a fraction of the cause and perpetuation of nonlinear effects. 
In fact, the network has the largest nonlinearity measure near $\alpha$ values with the most network induced nonlinearity (see Fig.~\ref{fig:nonlin_meas}). 
Since only a small portion of memristive edges are evolving actively in Fig.~\ref{fig:jnv_scatter}, the network appears to be underutilised; future studies (beyond simply changing the input layer) should optimise parameters which maximise the number of edges within this dynamically active region.

It has been identified that both network and memristive dynamics causes nonlinearity. 
To investigate its exact origin, a simple 3-node toy model was constructed, with node 0 serving as the drain (ground) node, node 1 as readout and node 2 as the input electrode node (see schematic in Fig.~\ref{fig:induced_flux_tunnel}), which takes in a sine wave (shifted away from zero) as input.
A highly nonlinear $x$ signal (extracted from the edge with the highest nonlinearity measure from Fig.~\ref{fig:jnv_scatter}) is artificially injected into the $1-2$ edge.
Both the modified and original $x_{1,2}$ values can be seen in Fig.~\ref{fig:induced_flux_tunnel}, where the modified $x_{1,2}$ display the classic behaviour of a memristive with high low-med dynamics measure as it never reaches the high conductance state. 
High--frequency fluctuations can only be observed in the readouts with the modified $x$ instance, confirming these nonlinear features are directly caused by these types of memristive edge dynamics. 

Performing the same analysis on the 3-node network with a binary memristor model yields vastly different results. 
As seen in Fig.~\ref{fig:induced_flux_binary}, despite having the same modified $x_{1,2}$ signal as Fig.~\ref{fig:induced_flux_tunnel}, the high--frequency fluctuations in the readouts are much more frequent, suggesting that the more physically realistic tunnelling memristor model may be more robust, and produce fewer high--frequency fluctuations when compared with the simpler binary model. 

Interestingly, the observed fluctuations are bounded between two values: the original sine wave readout and a constant zero signal (from the drain node).
This is due to conductance paths (specifically $g_{0,1}$ and $g_{1,2}$) rapidly switching in amplitude across time, favouring one signal over the other at different times; in the actual network the high--frequency fluctuations would be bounded by numerous other readouts in close proximity, as opposed to just two signals in this simple example. 
The gap at 34.5 Lyapunov times in Fig.~\ref{fig:induced_flux_binary} reveals how these fluctuations disappear when $|x_{1,2}|>x_{\rm crit}$, which explains the absence of high--frequency fluctuations with large low--high dynamics measures. 
Both Fig.~\ref{fig:induced_flux_tunnel} and \ref{fig:induced_flux_binary} shows examples of how nonlinear responses in $x_{1,2}$ can cause similar behaviours in the neighbouring $x_{0,1}$ values, illustrating the spread of similar memristive dynamics within the network which has potential to cause  observed nonlinear effects in the readouts. 

\section{Supporting Figures} \label{app:supp}

\begin{figure}
    \centering
    \includegraphics[width=0.8\linewidth]{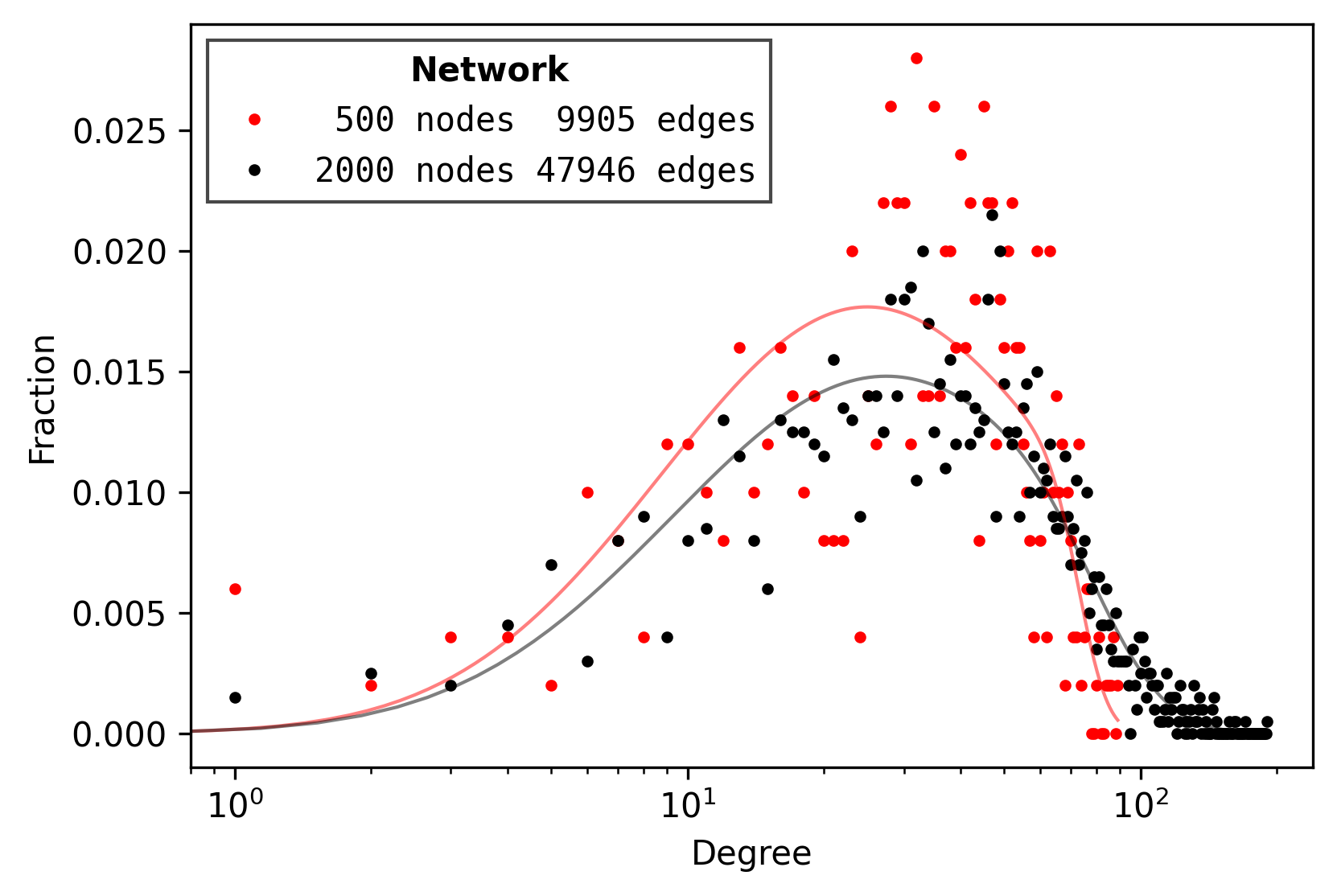}
    \caption{\rvb{Degree distribution of the two physical neuromorphic networks simulated in this study. Each marker represents the fraction (relative frequency) of each degree. The curves show the best fit skewed normal distribution, i.e. the networks have an approximately skewed log-normal degree distribution.
    For the $2,000$-node network, the shape, location and scale parameters of the best fit distribution are $-9.0$, $4.3$ and $1.0$ respectively. For the 500-node network, the parameters are $-3.7$, $4.5$ and $1.1$ respectively.}
    }
    \label{fig:degree_dist}
\end{figure}

\begin{figure}
	\includegraphics[width=\linewidth]{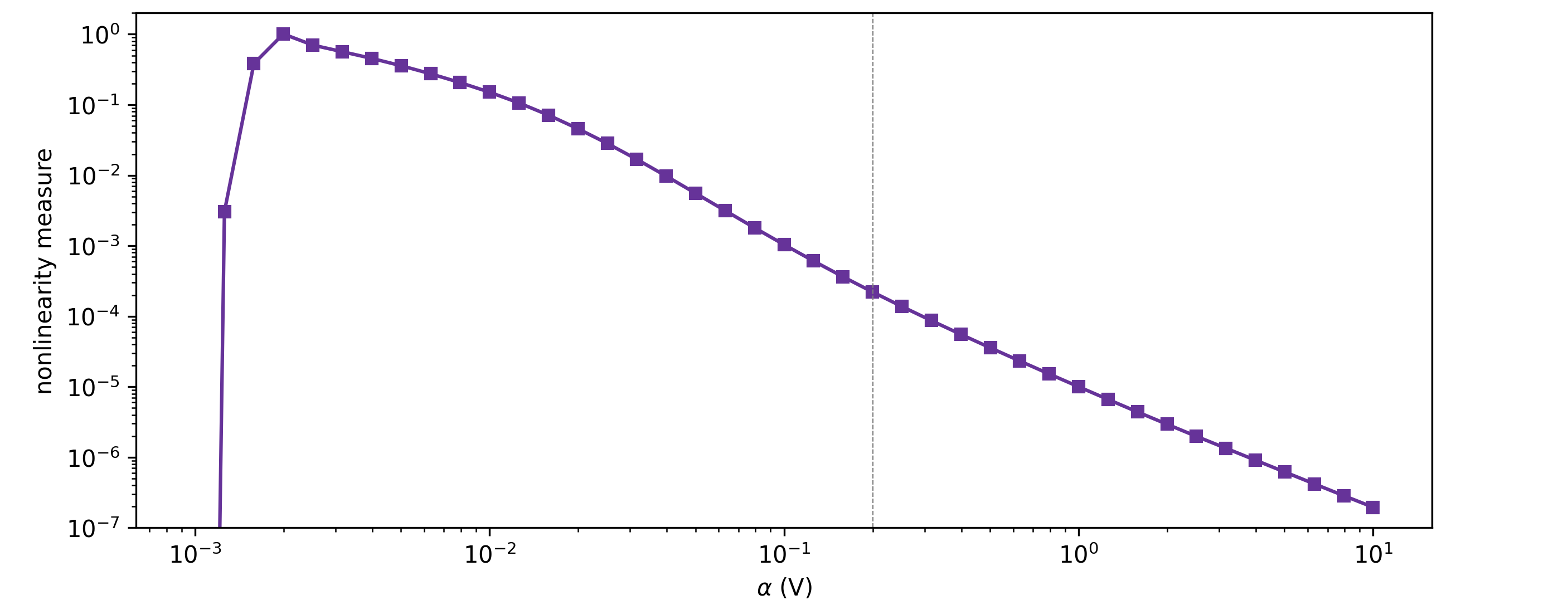}
	\caption{
		The average nonlinearity measure $\bar{m}_{\rm nonlin}$ (defined in Eq.~\eqref{eq:avg_m_nonlin}) as a function of the scaling parameter $\alpha$. 
		The simulations were performed on a neuromorphic network with $500$ nodes and $9,905$ edges. 
		The dotted line at $\alpha=0.2$ volts indicate the default $\alpha$ used. 
		Each data point was obtained from $100$ trials of the neuromorphic network simulation.
	}
	\label{fig:nonlin_meas}
\end{figure}

\begin{figure}
	\centering
	\includegraphics[width=\linewidth]{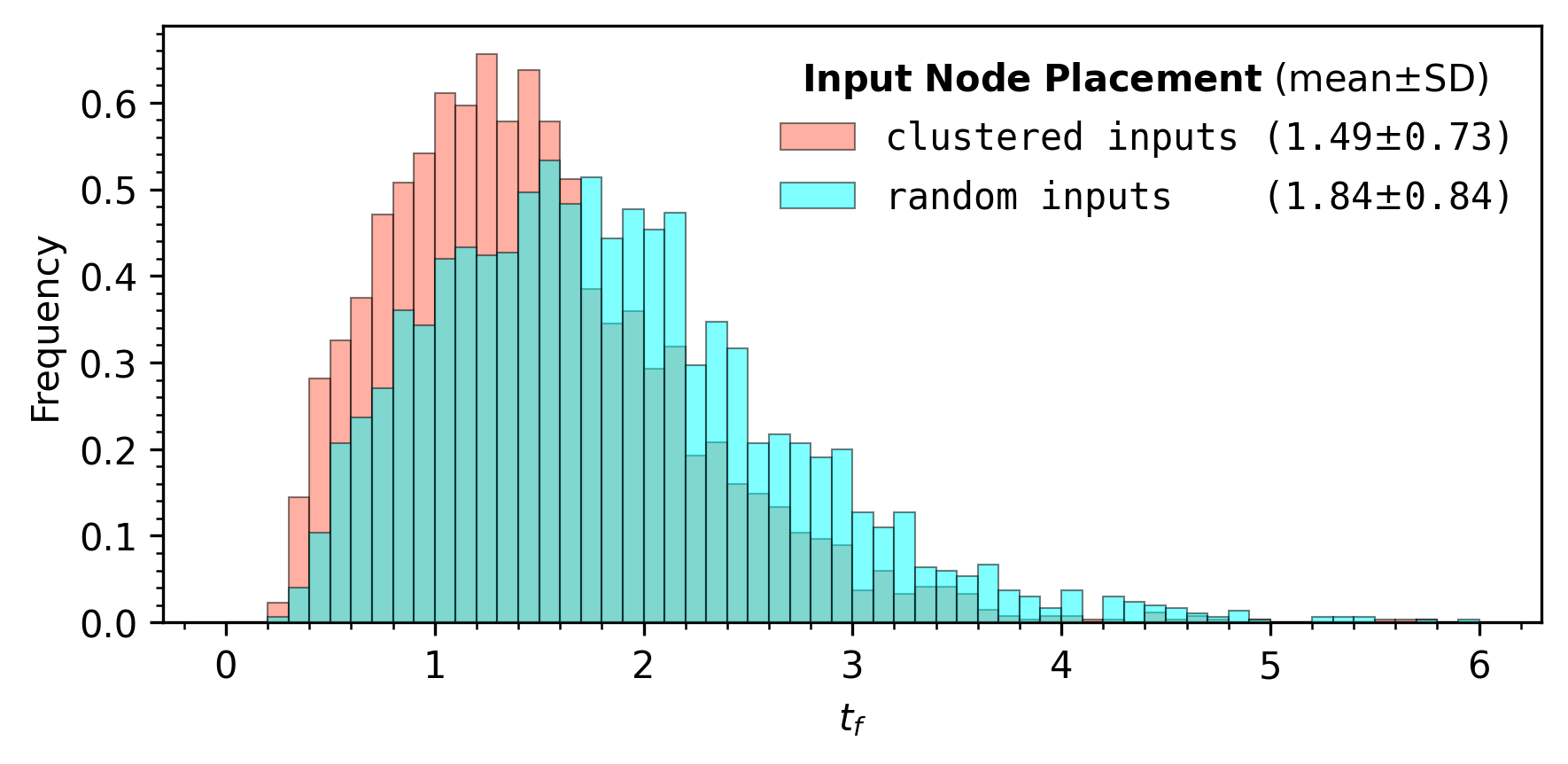}
	\caption{Effect of node-node distances for input node selection on forecast times. 
		Displayed are histograms of forecast times, with clustered inputs (each is within 2 distance of each other), and randomly selected inputs.
		Simulation performed on a neuromorphic network with 500 nodes and 9905 edges.}
	\label{fig:cluster_hist}
\end{figure}

\begin{figure}
	\centering
	\includegraphics[width=\linewidth]{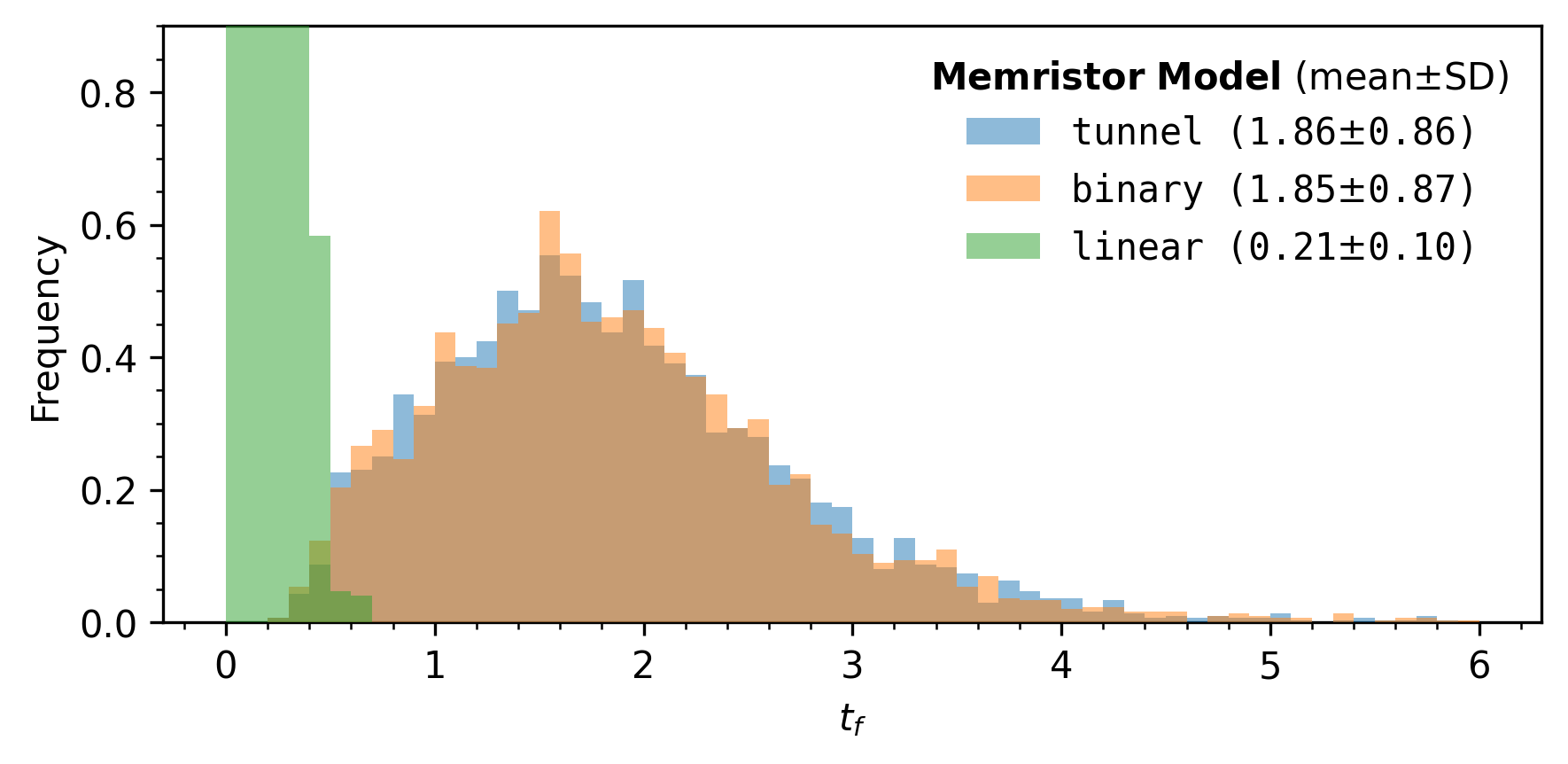}
	\caption{Effect of memristor model on forecast times. 
		Displayed are histograms of forecast times, with the (default) tunnelling, binary, and linear (i.e. resistor) models.
		Simulation performed on a neuromorphic network with $500$ nodes and 9905 edges, with $3,000$ data points per histogram.}
	\label{fig:bin_hist}
\end{figure}

\begin{figure}
	\centering
	\includegraphics[width=\linewidth]{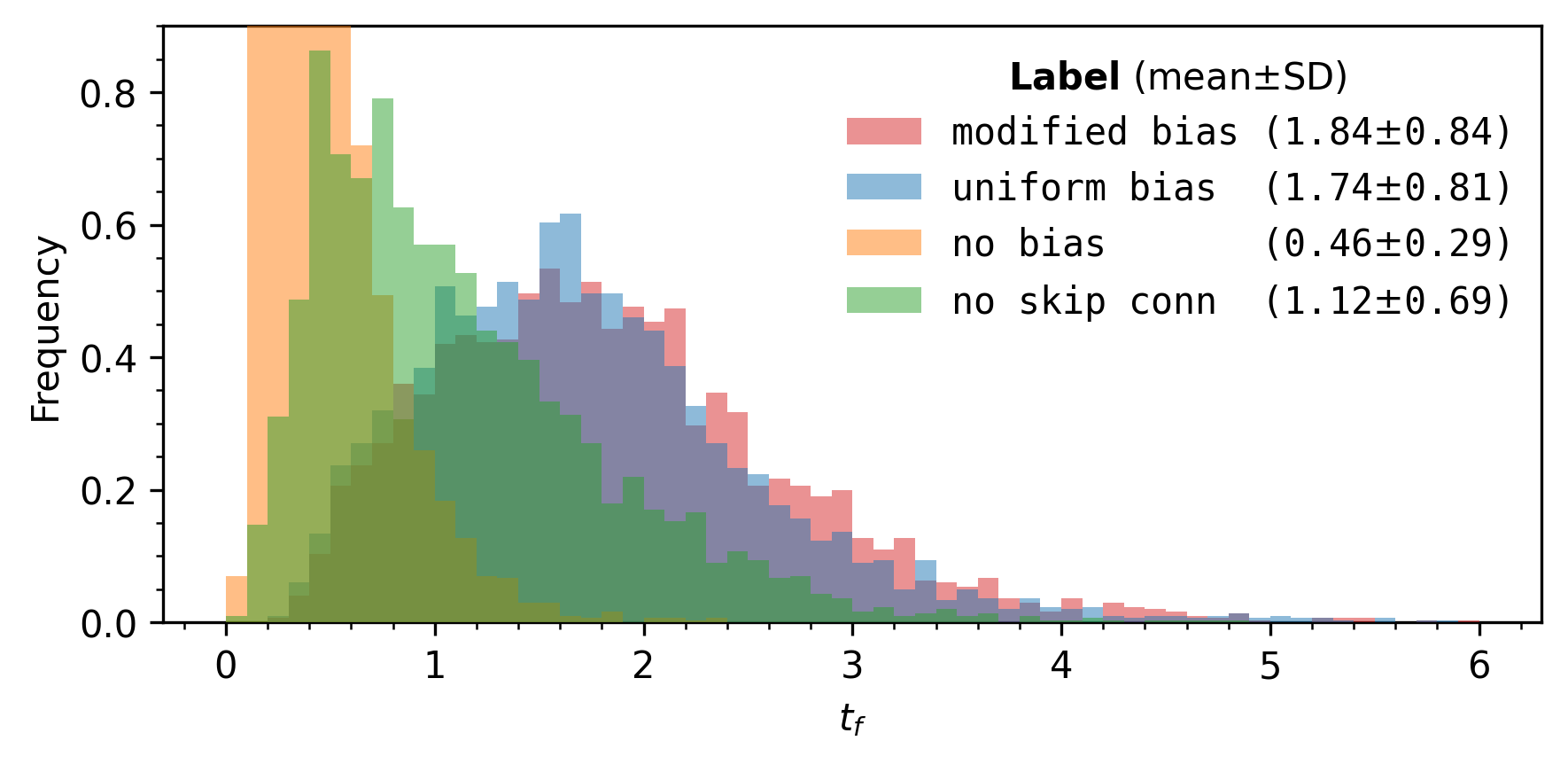}
	\caption{
		Effect of selected techniques (or lack thereof) used in this study on forecast times, displayed as histograms. 
		``Modified bias'' corresponds to the same parameters outlined in methods, with bias sampled from the interval $[-b,-b/2]\cup [b/2,b]$ .
		``Uniform bias'' has its values sampled from $[-b,b]$ (i.e. a uniform distribution $U(-b,b)$).
		``No bias'' is equivalent to $\mathbf{b}_{\rm in}=0$. 
		The skip connection is the additional $\mathbf{u}(t)$ term in Eq.~\eqref{eq:yhat}.
		Simulation performed on a neuromorphic network with $500$ nodes and $9,905$ edges, with $3,000$ data points per histogram.}
	\label{fig:ablation}
\end{figure}

Figs.~\ref{fig:degree_dist}--\ref{fig:ablation} are referred to in the main text.


\FloatBarrier
\bibliography{references}
\end{document}